\documentclass[twocolumn]{aastex62}

\usepackage{xspace}

\renewcommand{\rho}{\varrho}

\def \gray {$\gamma$-ray\xspace}
\def \grays {$\gamma$ rays\xspace}
\def \flx {photons $\mathrm{cm}^{-2}$ $\mathrm{s}^{-1}$}
\def \grid {AGILE-\textit{GRID}\xspace}

\def \lat {\textit{Fermi}-LAT\xspace}

\def \gcyg {G78.2+2.1\xspace}

\def \hpsr {PSR J2021+3651\xspace}
\def \gcygpsr {PSR J2021+4026\xspace}
\def \srcygpsr {PSR J2032+4127\xspace}

\received{...}
\revised{...}
\accepted{...}

\shorttitle{AGILE study of G78.2+2.1}
\shortauthors{Piano et al.}

\begin{document}

%\bibliographystyle{apj}
%\submitted{To be sumbmitted to \textit{ApJ Letters}. Draft version \today}

\title{AGILE study of the gamma-ray emission from the SNR G78.2+2.1 (Gamma Cygni)}

\correspondingauthor{Giovanni Piano}
\email{giovanni.piano@inaf.it}

\author{G. Piano}
\author{M. Cardillo}
\affiliation{INAF-IAPS, Via del Fosso del Cavaliere 100, I-00133, Roma, Italy}

\author{M. Pilia}
\author{A. Trois}
\affiliation{INAF-Osservatorio Astronomico di Cagliari, via della Scienza 5, I-09047, Selargius (CA), Italy}

\author{A. Giuliani}
\affiliation{INAF-IASF-Milano, via E. Bassini 15, I-20133 Milano, Italy}

\author{A. Bulgarelli}
\author{N. Parmiggiani}
\affiliation{INAF-OAS, Via P. Gobetti 93/3, I-40129 Bologna, Italy}

\author{M. Tavani}
\affiliation{INAF-IAPS, Via del Fosso del Cavaliere 100, I-00133, Roma, Italy}
\affiliation{INFN Roma Tor Vergata, Via della Ricerca Scientifica 1, I-00133, Roma, Italy}
\affiliation{Dipartimento di Fisica, Universit\`{a} di Roma ``Tor Vergata'', Via Orazio Raimondo 18, I-00173 Roma, Italy}

%\altaffiltext{1}{INAF/IAPS, via del Fosso del Cavaliere 100, I-00133 Roma, Italy}
%\altaffiltext{2}{CIFS-Torino, viale Settimio Severo 3, I-10133 Torino, Italy}
%\altaffiltext{3}{Dipartimento di Fisica, Universit\`a di Roma ``Tor Vergata'', via della Ricerca Scientifica 1,I-00133 Roma, Italy} 
%\altaffiltext{4}{INAF/IASF-Milano, via E. Bassini 15, I-20133 Milano, Italy}
%\altaffiltext{5}{INAF/IASF-Bologna, via Gobetti 101, 40129 Bologna, I-40129 Bologna, Italy}
%\altaffiltext{6}{Dipartimento di Fisica and INFN Trieste, via Valerio 2, I-34127 Trieste, Italy} 
%\altaffiltext{7}{INFN-Pavia, via Bassi 6, I-27100 Pavia, Italy}
%\altaffiltext{8}{INFN-Roma ``Tor Vergata'', via della Ricerca Scientifica 1, I-00133 Roma, Italy}
%\altaffiltext{9}{INAF-Osservatorio Astronomico di Cagliari, localit\`a Poggio dei Pini, strada 54, I-09012 Capoterra, Italy}
%\altaffiltext{10}{ENEA Frascati, via E. Fermi 45, I-00044 Frascati (Roma), Italy}
%\altaffiltext{11}{ASI Science Data Center, (ASDC), via G. Galilei, I-00044 Frascati (Roma), Italy}
%\altaffiltext{12}{INAF-OAR, I-00040, Via Frascati 33, Monte Porzio Catone, Italy} 
%\altaffiltext{13}{INAF-Osservatorio Astron. di Roma, Monte Porzio Catone, Italy}

\begin{abstract}
We present a study on the \gray emission detected by the \grid from the region of the SNR G78.2+2.1 (Gamma Cygni). In order to investigate the possible presence of \grays associated with the SNR below 1 GeV, it is necessary to analyze the \gray radiation underlying the strong emission from the pulsar \gcygpsr, which totally dominates the field. An ``off-pulse'' analysis has been carried out, by considering only the emission related to the pulsar off-pulse phase of the \grid light curve.  We found that the resulting off-pulsed emission in the region of the SNR -- detected by the \grid above 400 MeV -- partially overlaps the radio shell boundary.
By analyzing the averaged emission on the whole angular extent of the SNR, we found that a lepton-dominated double population scenario can account for the radio and \gray emission from the source. In particular, the MeV-GeV averaged emission can be explained mostly by Bremsstrahlung processes in a high density medium, whereas the GeV-TeV radiation by both Bremsstrahlung ($E_{\gamma} \lesssim 250$ GeV) and inverse Compton processes ($E_{\gamma} \gtrsim 250$ GeV) in a lower density medium.
\end{abstract}


\smallskip

%%%%%%%%%%%%%%%%%%%%%%%%%%%%%%%%%%%%%%%
\section{Introduction}
%%%%%%%%%%%%%%%%%%%%%%%%%%%%%%%%%%%%%%%

The radio source G78.2+2.1 (Gamma Cygni) is a typical shell-type Supernova Remnant (SNR) located within the extended emission of the Cygnus X region \citep{piddington_52}.\\
During the 70s, observations in the Cygnus X direction showed the presence of non-thermal emission from this source \citep{wendker_70,higgs_77a} and then \citet{higgs_77b}, with a study based on arcmin resolution observations at 1.42 GHz with the DRAO radiotelescope, confirmed the nature of this object: a SNR, with a shell structure of $\sim62'$ diameter, located at $(l,~b)=(78.2,~2.1)$. The inferred distance of the SNR is $d=1.5$ kpc $\pm$ 30\% \citep{landecker_80} and the corresponding shell radius is $R\simeq13.5$ pc.  By analyzing ASCA X-ray data, \citet{uchiyama_02} observed a region (R3) with an arclike morphology along the outer boundary of the radio spherical shell emitting thermal radiation with $kT_e = 0.76^{+0.10}_{-0.09}$ keV. This feature indicates the presence of a thin thermal plasma in the SNR post-shock region propagating through the interstellar medium (ISM). Assuming a strong shock scenario, they inferred a shock velocity $v_s = 800^{+50}_{-60}$ $\mathrm{km~s^{-1}}$ in the light  of the relation with the electron temperature. Provided that the SNR is in the Sedov adiabatic expansion phase, they estimated the age of G78.2+2.1, $\tau_{age}\simeq6600$ yr. Moreover, they reported the discovery of some clumps (in the North-Western part of the SNR) with a hard power-law tail ($\Gamma \sim 0.8 - 1.5$) spectral component (3--8 keV), possibly related to non-thermal Bremsstrahlung and indicating a shock-cloud interaction.

\citet{ladouceur_08} studied this SNR very extensively, by analyzing radio continuum emission at 408 and 1420 MHz, radio 21 cm line emission of HI, infrared continuum emission at 8.23 $\mu$m and 60 $\mu$m. They found that the SNR non-thermal synchrotron emission has a quasi-perfect circular shape remarking its spherical symmetry.
Furthermore, they analyzed the HI kinematics leading to a model of the SNR dynamical evolution through the surrounding gas. The most probable scenario suggests that the SNR progenitor would have largely evacuated a cavity around the star through a strong stellar wind. An expanding shell of HI has been set in motion by this wind with an expanding velocity of few $\mathrm{km~s^{-1}}$.

The first \gray source ever associated to the SNR G78.2+2.1 was 2CG 078+2, detected by the COS B satellite \citep{pollock_85a}. They detected a \gray source centered at $(l,~b)=(78.1,~2.3)$ with a photon flux above 300 MeV of about $F_{\gamma} \approx 4.8 \cdot 10^{-8}$ \flx. \citet{pollock_85b} proposed a probable association with the SNR, based on the spatial coincidence of the \gray source with a bright radio feature (DR4, \citealp{downes_66}) and a molecular cloud near the remnant (source number 8 in the CO survey of the Cygnus X region by \citealp{cong_77}). According to the author's interpretation, $\gamma$ rays are produced by Bremsstrahlung emission from accelerated electrons in correspondence with molecular cloud/shock interaction.

Nevertheless, the association with the SNR remained doubtful. In fact, the EGRET associated \gray source, 3EG J2020+4017, was presented as the brightest unidentified \textit{point source} in the 3rd EGRET Catalog \citep{egretcat}. It is located at $(l,~b)=(78.05,~2.08)$, with a 95\% C.L. error box of $0.16^{\circ}$ and a photon flux above 100 MeV of $F_{\gamma} = (123.7 \pm 6.7) \cdot 10^{-8}$ \flx.

\begin{figure}[!ht]
 \begin{center}
  \includegraphics[width=8.5cm]{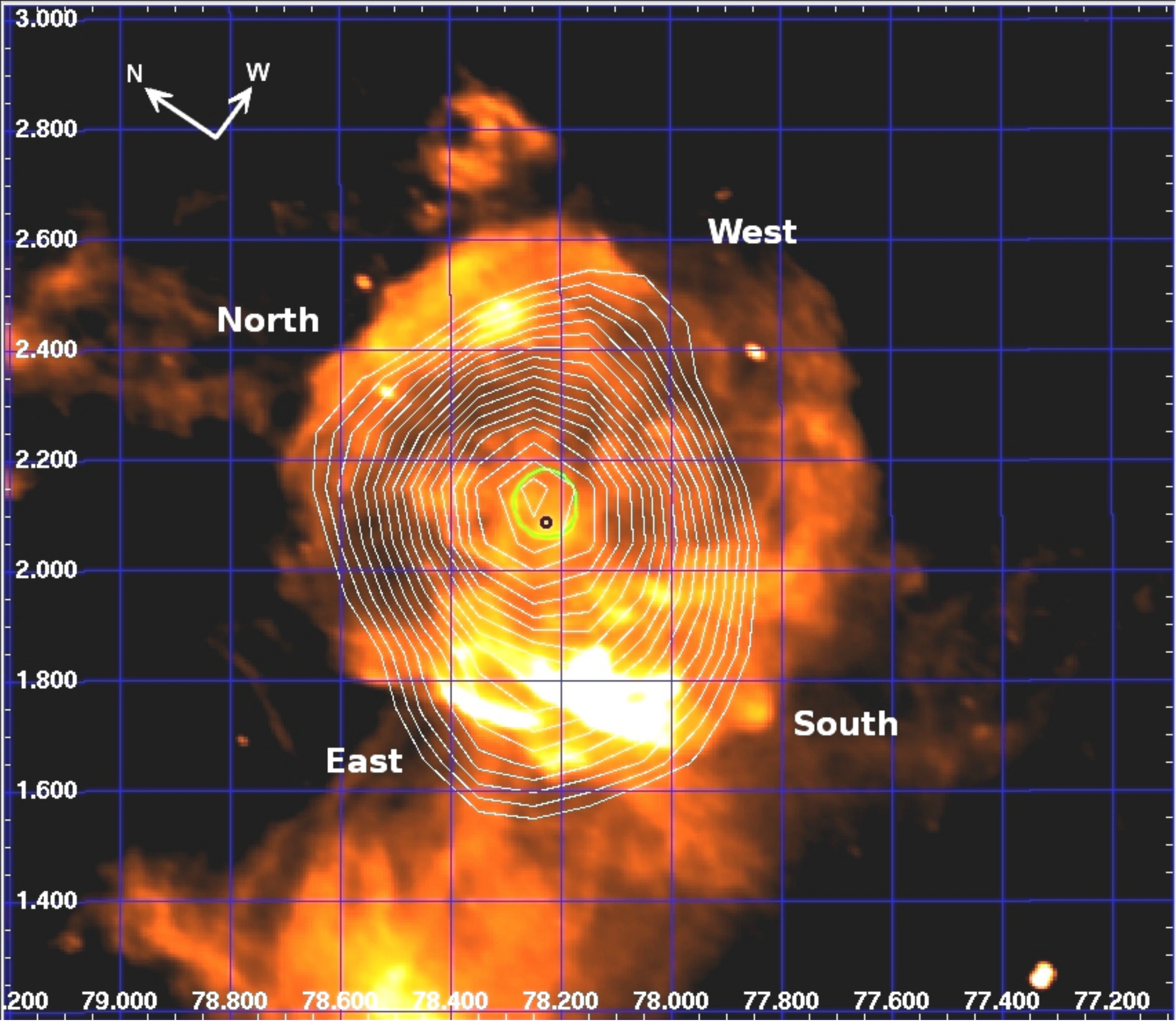}
 \caption{Gamma Cygni SNR (G78.2+2.1) in Galactic coordinates. Image obtained with the Dominion Radio Astrophysical Observatory (DRAO), from the Canadian Galactic Plane Survey (CGPS), wavelength = 21.1 cm (frequency = 1420 MHz, bandwidth = 30 MHz), pixel size = 20$''$ \citep{taylor_03}. \textit{White} contour levels: \grid intensity contour levels above 100 MeV -- related to data integrated between 2007-November-02 and 2009-July-29 -- (pixel size $0.1^{\circ}$), starting from 0.00085 in steps of 0.00002 (intensity per pixel); \textit{green} contour: \grid 95\% confidence level for $E \ge 100~$MeV; \textit{black} circle: \gcygpsr. Orientation of the Celestial coordinates is marked.}
 \label{gamma_cyg_snr}
 \end{center}
 \end{figure}

The \grid (Astrorivelatore Gamma ad Immagini LEggero - Gamma Ray Imaging Detector, \citealp{tavani_09,barbiellini_02,prest_03}) observations found that the corresponding source, 1AGL J2022+4032 \citep{agilecat}, has a position and a \gray photon flux above 100 MeV consistent with the EGRET values. We found a bright point source within a widespread \gray emission morphologically consistent with the circular shape detected at radio wavelengths (see Fig.~\ref{gamma_cyg_snr}). 1AGL J2022+4032 is a puzzling \gray source, showing some hints of variability and it is likely to be a superposition of multiple point-like sources emitting in \grays (see \citealp{chen_piano_11}).

\lat \citep{atwood_09} has identified this source as the bright \gray pulsar \gcygpsr \citep[3FGL J2021.5+4026,][]{abdo_09,abdo_10a,abdo_10b,abdo_13,acero_15}, characterized by a sharp exponential cut-off at $E = (2.37 \pm 0.06$) GeV. The \lat observations found a significant decrease in flux (for $E > 100$ MeV) from this pulsar around mid-October 2011, with a simultaneous increase in the frequency spindown rate \citep{allafort_13}. The X-ray counterpart of \gcygpsr is 2XMM J202131.0+402645 \citep{trepl_10, weisskopf11}, whose pulsations were discovered in the X-ray band and reported by \citet{lin_13}. No radio pulsations have been detected so far from this source.

Above 10 GeV, \lat detected a significant extended source that is consistent with the spherical shell of the SNR detected at radio wavelengths \citep{lande_12,ackermann_17}. Furthermore, a bright \gray excess is detected in the North-Western (NW) rim of the remnant at GeV (\lat, \citealp{fraija_16}) and TeV energies (VERITAS, \citealp{aliu_13, abeysekara_18}), VER J2019+407 (see Fig.~\ref{gamma_cyg_radio_TeV}). SNR G78.2+2.1 was included in the first \lat SNR catalog, a survey ranging from 1 to 100 GeV \citep{acero_16}.

Therefore, in this region the overall \gray emission between 100 MeV and a few GeV is totally dominated by the strong pulsar, with the high-energy emission from the SNR only accounting for an underlying component of the field. In order to evaluate the presence of diffuse high-energy emission correlated to the SNR, we analyzed the emission in the field within the off-pulse interval of the bright pulsar \gcygpsr.

%%%%%%%%%%%%%%%%%%%%%%%%%%%%%%%%%%%%%%%
\section{Data Analysis}
%%%%%%%%%%%%%%%%%%%%%%%%%%%%%%%%%%%%%%%

The AGILE satellite was launched on April 23, 2007. Until October 2009 it worked in ``pointing'' mode with fixed attitude, and from November 2009 it operates in ``scanning mode'', with a controlled rotation of the pointing axis.
During the ``pointing'' mode data-taking ($\sim$2.5 years) the AGILE satellite performed $\sim$100 pointings, called Observation Blocks (OBs) with variable exposure times (typically 3--30 days), drifting about 1 degree per day with respect to the initial boresight direction to match solar panels illumination constraints \footnote{A detailed schedule of the AGILE Observation Block is available online at \htmladdnormallink{http://agile.asdc.asi.it/current\_pointing.html}{http://agile.asdc.asi.it/current\_pointing.html}}. In this configuration the \grid was characterized by enhanced performances in the monitoring capability of a given source, especially in the energy band 100--400 MeV (see \citealp{bulgarelli_12} for details). The Cygnus region was one of the preferred target of the AGILE pointing plan. 	Thus, we were able to obtain a deep exposure of this sky region already in the relatively short pointing phase.

In this paper we present an analysis of this region on a deep integration of the \grid data from 2007-November-02 to 2009-July-29 (the same dataset reported by \citealp{chen_piano_11, piano_12}). During this period AGILE repeatedly pointed at the Cygnus region for a total of $\sim$275 days, corresponding to a net exposure time of $\sim$11 Ms.
The analysis was carried out by using the last available \grid software package (Build 25), \verb+FM3.119+ calibrated filter, \verb+I0025+ response matrices, and consolidated archive (\verb+ASDCe+) from ASI Data Center.

In order to investigate the presence of $\gamma$ rays produced by the SNR shock, we tried to unveil the emission underlying the pulsar dominating radiation, by performing an ``off-pulse'' analysis \citep{pellizzoni10}.

The pulsar ephemeris we used was obtained from the \lat data. Because \gcygpsr is known to be variable \citep{allafort_13}, we selected an interval as close as possible to our observations (June 2008 to May 2009) and well before the ``jump'' which was observed by \cite{allafort_13} in October 2011.
Given the AGILE pointing observing strategy, further addition of pointing data did not provide a significant improvement in the statistics and  significance of the detection. Thus, we decided to keep the same interval as in \cite{chen_piano_11} as for consistency. 
We decided to not extend the solution to the scanning mode dataset, because even if it keeps the imaging accuracy \citep{piano_2017,piano_2018}, it does not improve the analysis of the timing profile. This is probably due to instabilities in the pulsating emission of the source during this period.

\begin{figure}[!ht]
 \begin{center}
  \includegraphics[width=8.0cm]{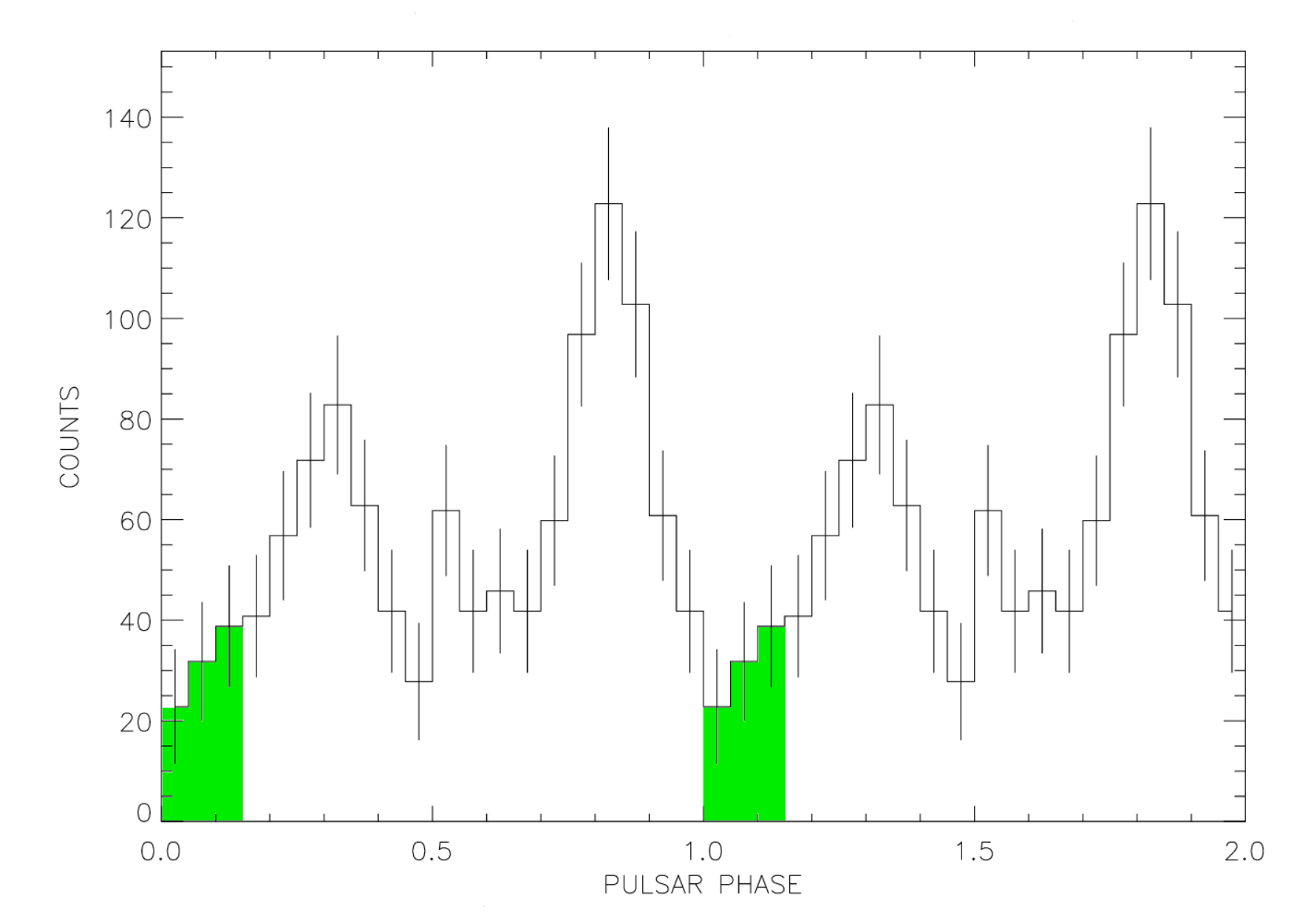}
\caption{\grid folded light curve of \gcygpsr, photons with $E>100$ MeV, 2 cycles with 20 bins per cycle. The green regions highlight the off-pulse phase, $0.00 \leqslant \Delta\phi_{off} \leqslant 0.15$, adopted in this analysis.}
\label{psr_gamma_cyg_off_pulse}
 \end{center}
 \end{figure}

Based on the pulsed signal analysis described in \cite{pellizzoni09}, we selected only high confidence \gray photons (G) with energy above 100 MeV within a radius of 1 degree from the pulsar position (R.A.: $20^h:21^m:30^s.733$, Dec.: $+40^{\circ}:26':46''.04$) (J2000, \citealt{weisskopf11}) so as to avoid contamination from the nearby crowded environment. We performed a first folding over the nominal period and period derivative ($P = 0.26531701593(3)$\,s, $\dot P = 5.483(3) \times 10^{-14}$), producing a detection at 4$\sigma$. Because the ephemeris did not fully cover the data interval, we also chose to perform a 5$\sigma$ search around the nominal period in order to fully exploit AGILE's temporal resolution resulting from the long observing time span. We found that the best period for this interval is $P = 0.26531701607(9)$\,s, with a weighted post-trial detection significance of 5.3$\sigma$.
The \grid light curve folded on the corresponding period of the \gcygpsr is shown in Fig.~\ref{psr_gamma_cyg_off_pulse}.

The pulsed profile of \gcygpsr is not the common double-peaked \gray profile of many energetic \gray pulsars: it is characterized by a high unpulsed fraction and a not sharp separation between on-peak and off-peak phases. Thus, the \grid off-pulse profile for this object was quite complex to determine. We decided to adopt, as the best-defined off-pulse phase, the phase interval $0.00 \leqslant \Delta\phi \leqslant 0.15$ (green-highlighted region in Fig.~\ref{psr_gamma_cyg_off_pulse}).

\begin{figure}[!t]
 \begin{center}
 \begin{tabular}{cc}
  \includegraphics[width=4.0cm]{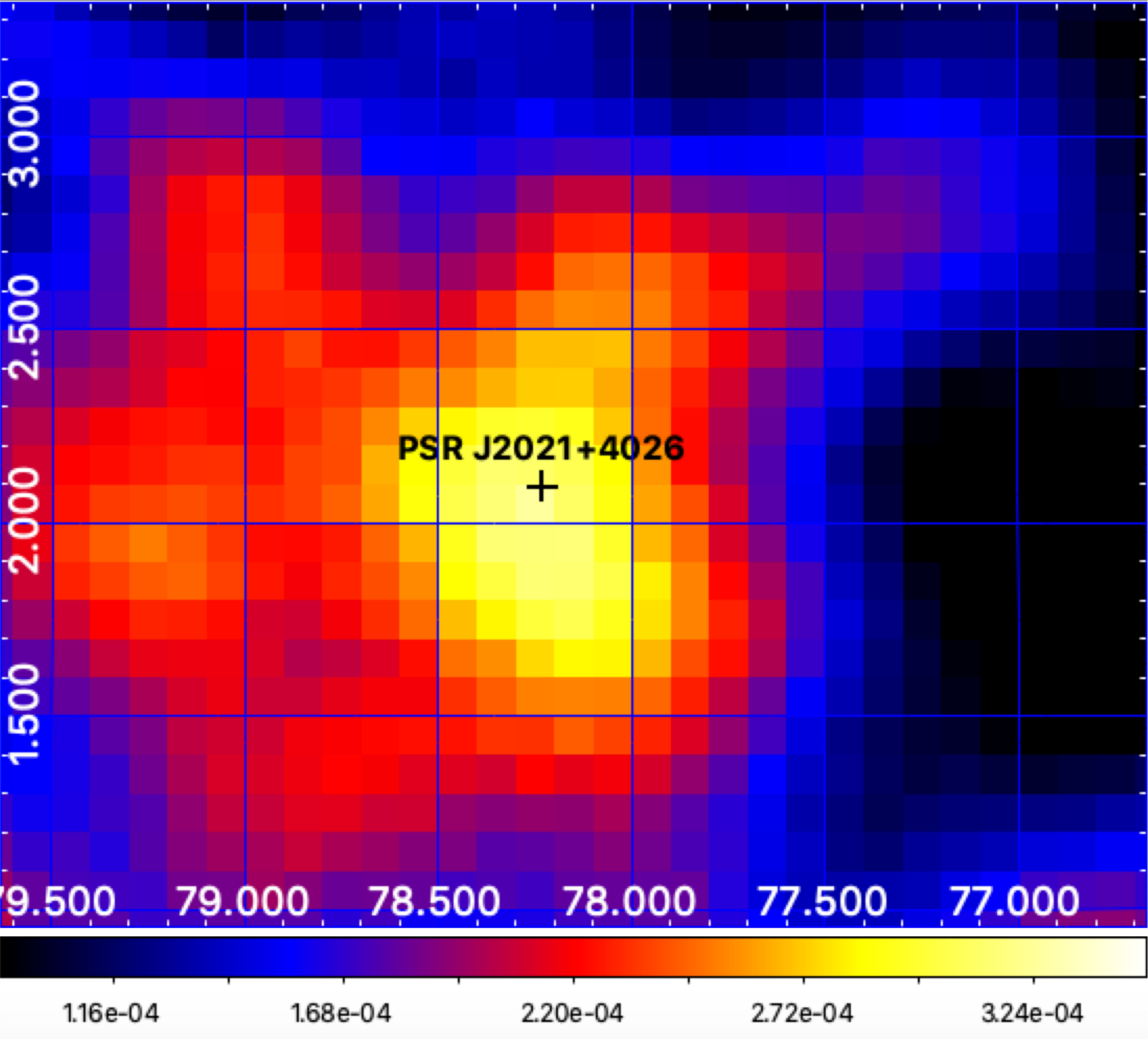} & \includegraphics[width=4.0cm]{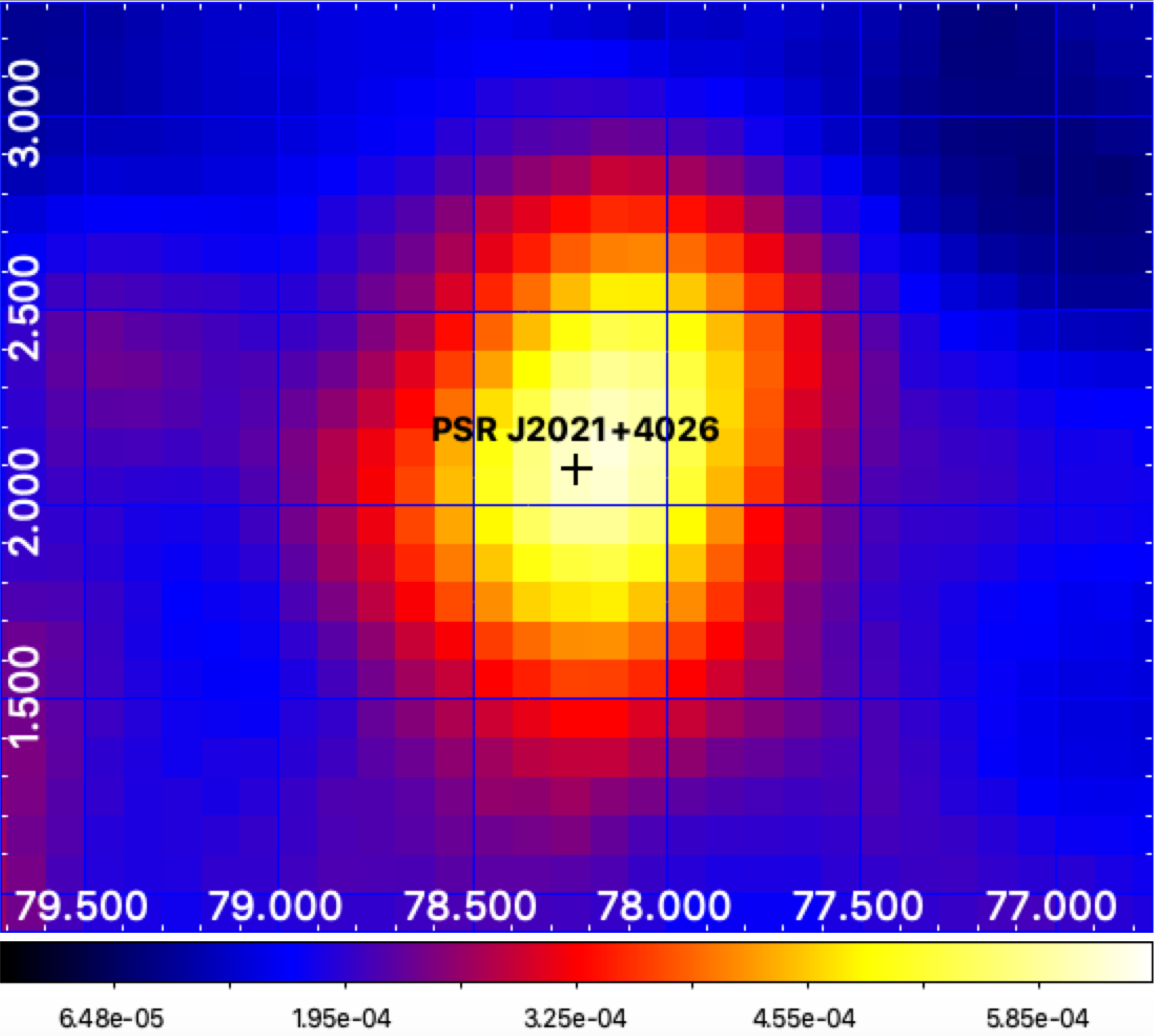}%  
 \end{tabular}
 \caption{\grid intensity maps above 400 MeV. Pixel size = $0.1^{\circ}$ with 5-pixel Gaussian smoothing, color bar scale in units of $\mathrm{photons~cm^{-2}~s^{-1}~pixel^{-1}}$.  The \textit{black crosses} mark the position of the pulsar \gcygpsr. \textit{Left panel}: intensity map related to the off-pulse interval ($0.00 \leqslant \Delta\phi \leqslant 0.15$, green band in Fig.~\ref{psr_gamma_cyg_off_pulse}). \textit{Right panel}: intensity map related to the phase interval $0.25 \leqslant \Delta\phi \leqslant 0.40$.}
 \label{gamma_cyg_phases}
 \end{center}
 \end{figure}

In Fig.~\ref{gamma_cyg_phases} the \grid intensity map related to the off-pulse interval ($0.00 \leqslant \Delta\phi \leqslant 0.15$, green band in Fig.~\ref{psr_gamma_cyg_off_pulse}) is compared to the map related to the secondary peak interval ($0.25 \leqslant \Delta\phi \leqslant 0.40$). We note that, whereas in the latter, the pulsar is still bright and dominates the field, in the former, a weaker pulsar remnant is detected together with a diffuse emission in the South-Eastern (SE) rim of the shell. The off-pulse image above 400 MeV shows a complex extended-source morphology.

\begin{figure}[!h]
 \begin{center}
  \includegraphics[width=8.0cm]{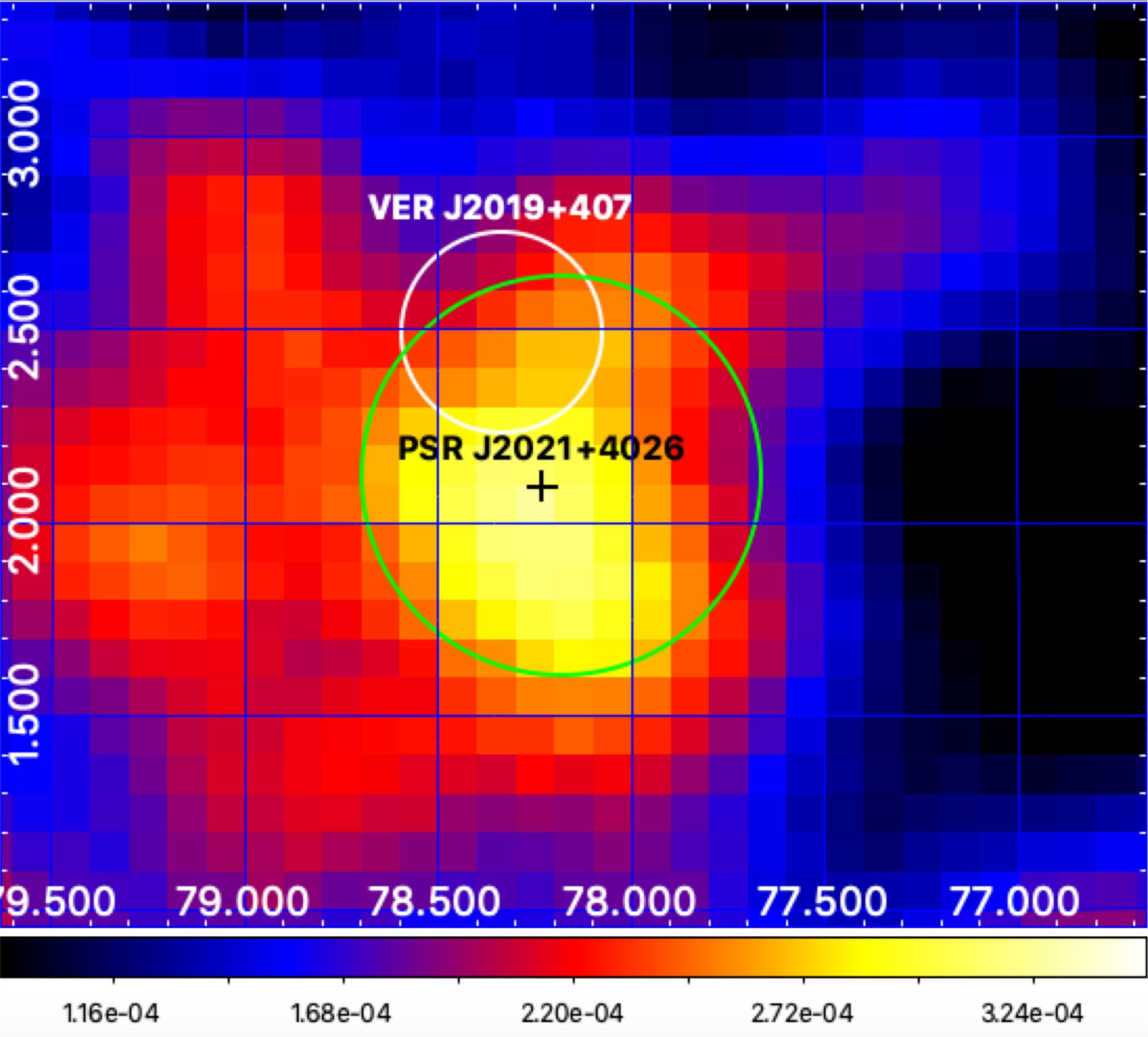}
 \caption{\grid off-pulse (phase: 0.00-0.15) intensity map above 400 MeV of the Gamma Cygni SNR. Pixel size = $0.1^{\circ}$ with 5-pixel Gaussian smoothing, color bar scale in units of $\mathrm{photons~cm^{-2}~s^{-1}~pixel^{-1}}$.  The \textit{black cross} marks the position of the pulsar \gcygpsr. The \textit{green circle} marks the extent of the SNR synchrotron shell as visible at radio wavelengths. The \textit{white circle} marks the position of the extended TeV source, VER J2019+407, as detected by VERITAS \citep{aliu_13}.}
\label{gamma_cyg_radio_TeV}
 \end{center}
 \end{figure}

The imaging of the SNR region has been analyzed above 400 MeV (Fig.~\ref{gamma_cyg_radio_TeV}), considering the \gray emission weighted on the residual 15\% exposure time. We decided to investigate the imaging profile above 400 MeV (and not above 100 MeV) because of a better \grid point spread function (PSF) at those energies ($4.2^{\circ}$ at 100 MeV, $1.2^{\circ}$ at 400 MeV, 68\% containment radius, \citealp{sabatini_15}). In doing so, we can better investigate the correlation with other wavelengths.

\begin{figure}[!h]
 \begin{center}
  \includegraphics[width=8.0cm]{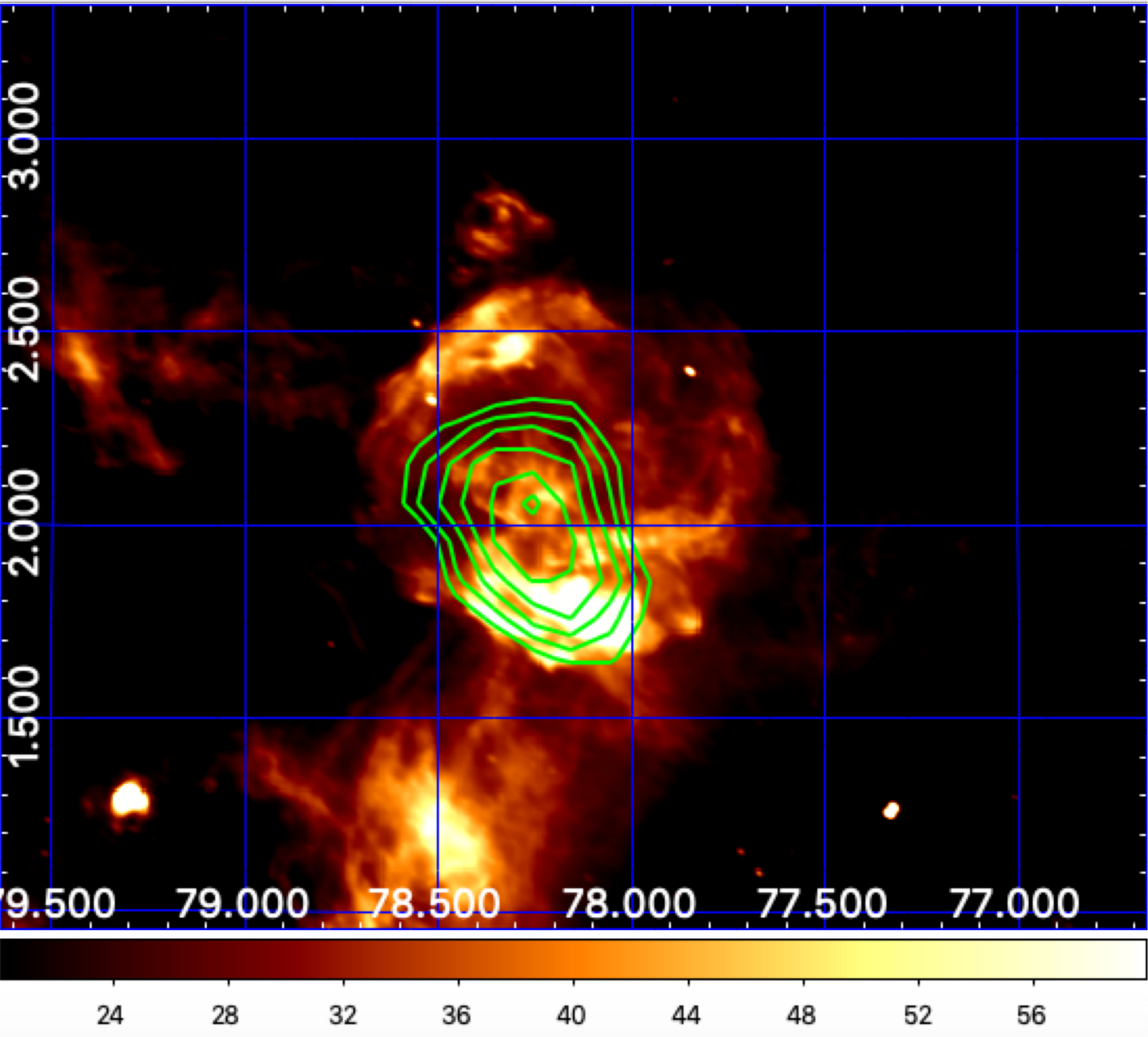}
 \caption{Radio image of the SNR G78.2+2.1 at 21.1 cm wavelength, (frequency = 1420 MHz, bandwidth = 30 MHz), pixel size = 20$''$, DRAO Radio Telescope. Overlaid are the \grid contour levels (\textit{green}), related to the off-pulse intensity map above 400 MeV.}
 \label{gamma_cyg_radio_agilecon}
 \end{center}
 \end{figure}

In Fig.~\ref{gamma_cyg_radio_agilecon} the \grid contours of the off-pulse intensity map above 400 MeV are superimposed on the radio map (wavelength = 21.1 cm) of the SNR.
If the emission underlying the intense \gray radiation from the pulsar is unveiled, an extended and complex \gray morphology comes out, presumably correlated to the SNR.
The \gray emission shape within the SNR shell partially covers the boundary of the remnant in coincidence with a bright SE feature (DR4, \citealp{downes_66}) of the radio synchrotron shell (see the \grid contour levels in Fig.~\ref{gamma_cyg_radio_agilecon}). 

Finally an extended-source MSLA (multi-source likelihood analysis, \verb+multi+ version of the \grid software package) was carried out on the off-pulse map above 400 MeV, taking into account the emission from the three Cygnus \gray pulsars (\gcygpsr, \srcygpsr and \hpsr). In order to account for the reduced exposure, the photon flux values for these persistent objects have been set as free parameters in the multi-source analysis and re-estimated. Firstly, we estimated the photon fluxes of the distant pulsars: \hpsr and \srcygpsr. Secondly, we kept the spectral parameters fixed for those pulsars to the ones found in the first step and analyzed together the SNR with the pulsar \gcygpsr. Our MSLA software \citep{bulgarelli_12} is based on the Test Statistic (TS) method as formulated by \citet{mattox_96}. This statistical analysis provides a detection significance assessment of a \gray source by comparing maximum-likelihood values of the null hypothesis (no source in the model) with the alternative hypothesis (point or extended source in the field model).

\begin{figure*}[!t]
 \begin{center}
 \begin{tabular}{cc}
  \includegraphics[width=8.0cm]{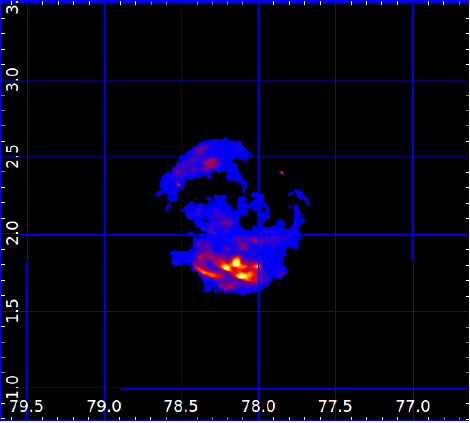} & \includegraphics[width=8.0cm]{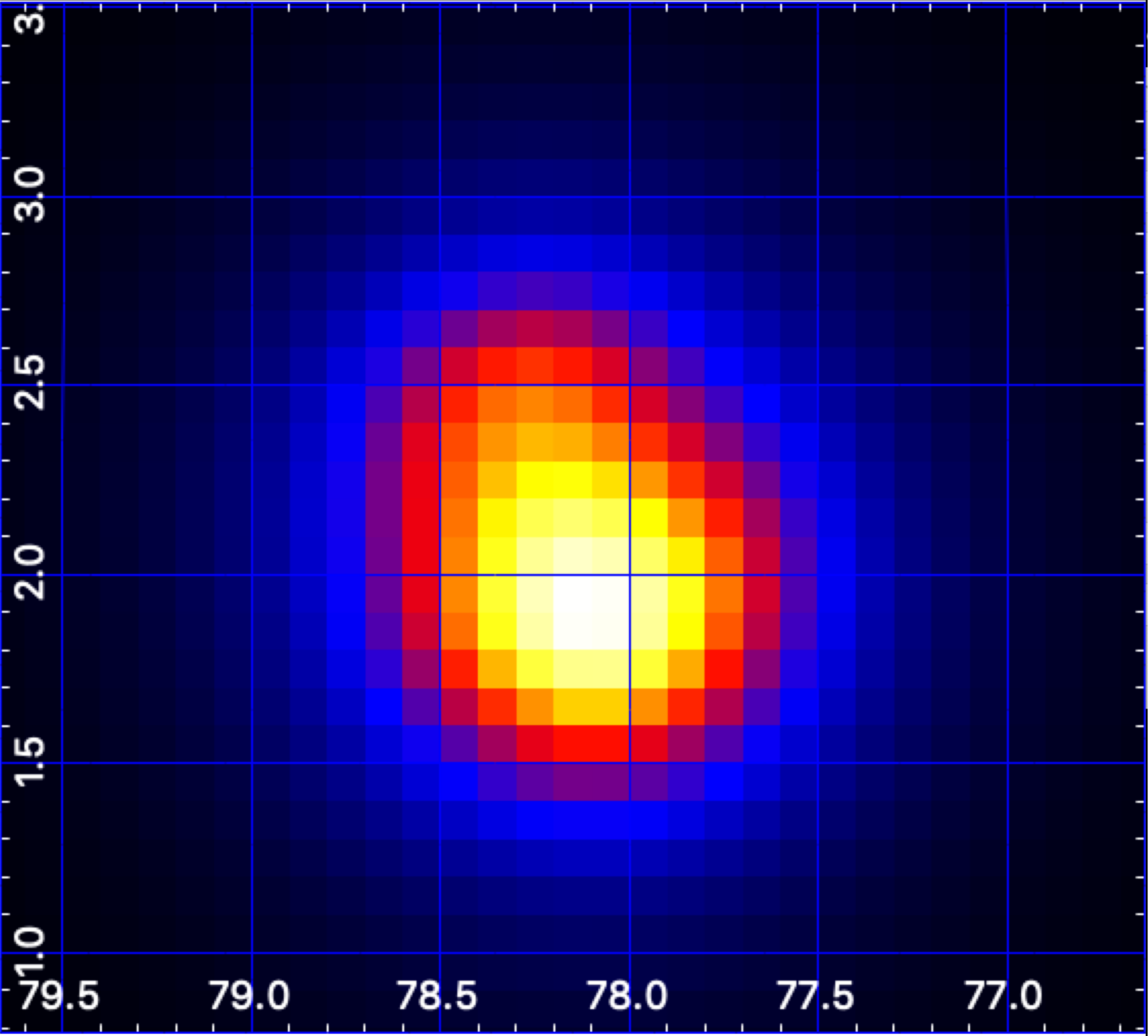}%  
 \end{tabular}
 \caption{\textit{Left panel}: radio emission related to the quasi-symmetrical synchrotron shell in Galactic coordinates. \textit{Right panel}: the same map convolved with the \grid PSF above 400 MeV, and used as template for the extended-source MSLA. The convolved template has a 68\% containment radius of $\sim$1.0 degrees.}
 \label{gamma_cyg_radio_template}
 \end{center}
 \end{figure*}

In order to perform an extended-source analysis, we adopt the radio quasi-symmetrical shell properly convolved with the \grid PSF above 400 MeV as template for the maximum likelihood study (see Fig.~\ref{gamma_cyg_radio_template}). We found a detection significance of $\sqrt{TS} = 8.6$ and a flux $f (E > 400 \ \mathrm{MeV}) = (14.8 \pm 2.2) \ 10^{-8} \ \mathrm{photons~cm^{-2}s^{-1}}$. The remnant of the \gcygpsr is weakly detected ($\sqrt{TS} \simeq 2$) with a flux $f (E > 400 \ \mathrm{MeV}) = (7.1 \pm 3.2) \ 10^{-8} \ \mathrm{photons~cm^{-2}s^{-1}}$.

\begin{figure}[!h]
 \begin{center}
  \includegraphics[width=9.0cm]{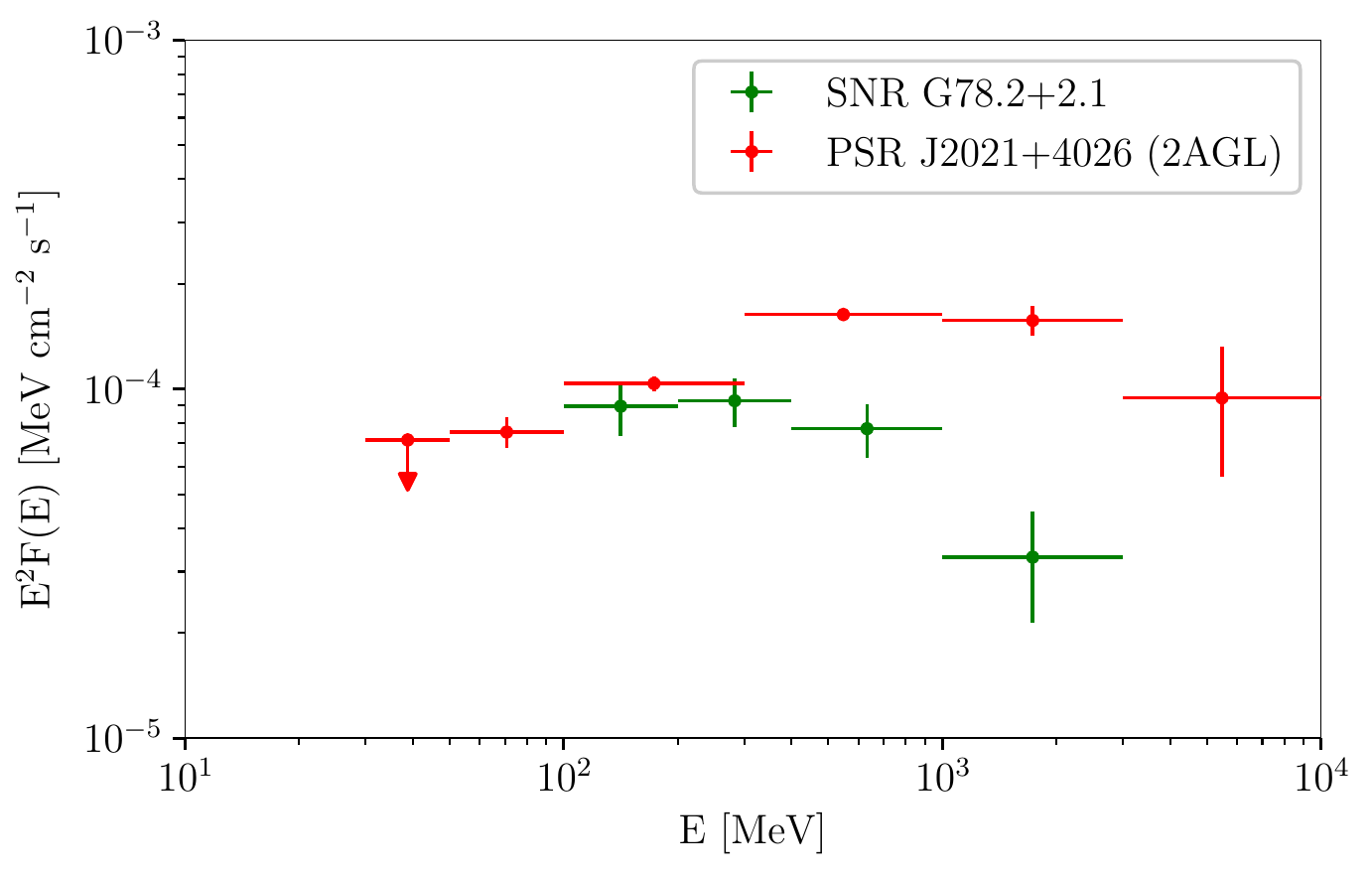}
 \caption{Spectrum of the SNR G78.2+2.1 as detected with the off-pulse analysis by AGILE (\textit{green}), compared with the spectrum of the pulsar \gcygpsr from the second AGILE catalog (\textit{red},  2AGL J2021+4029, \citealp{bulgarelli_19}).}
 \label{gamma_cyg_2AGL_sed}
 \end{center}
 \end{figure}

Furthermore, we carried out a spectral analysis between 100 MeV and 3 GeV, still assuming the same extended-source template, convolved with the corresponding PSF of each energy band. The results of this analysis are shown in Table \ref{tab:spectrum} and Fig.~\ref{gamma_cyg_2AGL_sed}, where the spectrum associated with the SNR (as calculated in this paper) is compared with the AGILE spectrum of the pulsar \gcygpsr (2AGL J2021+4029, as presented in the Second AGILE catalog, \citealp{bulgarelli_19}). We note that the pulsar is characterized by a harder spectrum with respect to the SNR. From the folded light curve of Fig.~\ref{psr_gamma_cyg_off_pulse}, we expect a sub-dominant contamination of the pulsar remnant in the off-pulse dataset.

\begin{table*}
	\begin{center}
	\caption{Spectrum of the SNR G78.2+2.1, as detected by AGILE: energy band, significance of detection ($\sqrt{TS}$), flux, radius of the SNR template convolved with the AGILE PSF (68\% containment radius).}
 {\small
 		\begin{tabular}{c|c|c|c}
 		\label{tab:spectrum}
 Energy band & $\sqrt{TS}$ &                                   flux                                        & template radius \\
     (MeV)      &                    & ($10^{-8} \ \mathrm{photons \ cm^{-2} \ s^{-1}}$) &       (degrees)    \\
 \hline\hline
  100-200     &        6.14     &                            44.9 $\pm$ 7.9                            &      $\sim$3.5   \\
  200-400     &        7.22     &                            23.3 $\pm$ 3.7                            &      $\sim$1.9    \\
  400-1000   &        7.07     &                            11.6 $\pm$ 2.1                            &      $\sim$1.1     \\
1000-3000   &        3.80     &                              2.2 $\pm$ 0.8                            &      $\sim$0.7     \\
		\end{tabular}
 }
 \end{center}
\end{table*}

%%%%%%%%%%%%%%%%%%%%%%%%%%%%%%%%%%%%%%%
\section{Multiwavelength imaging}
%%%%%%%%%%%%%%%%%%%%%%%%%%%%%%%%%%%%%%%

In this Section we present the multiwavelength emission in the SNR region. First, we considered the gas distribution, being the target of the accelerated high-energy particles and then we evaluated the pattern of the high-energy emission, from X-rays to VHE \grays.

\subsection{Gas distribution}
\begin{figure*}[!t]
 \begin{center}
 \begin{tabular}{ccc}
  \includegraphics[width=5.8cm]{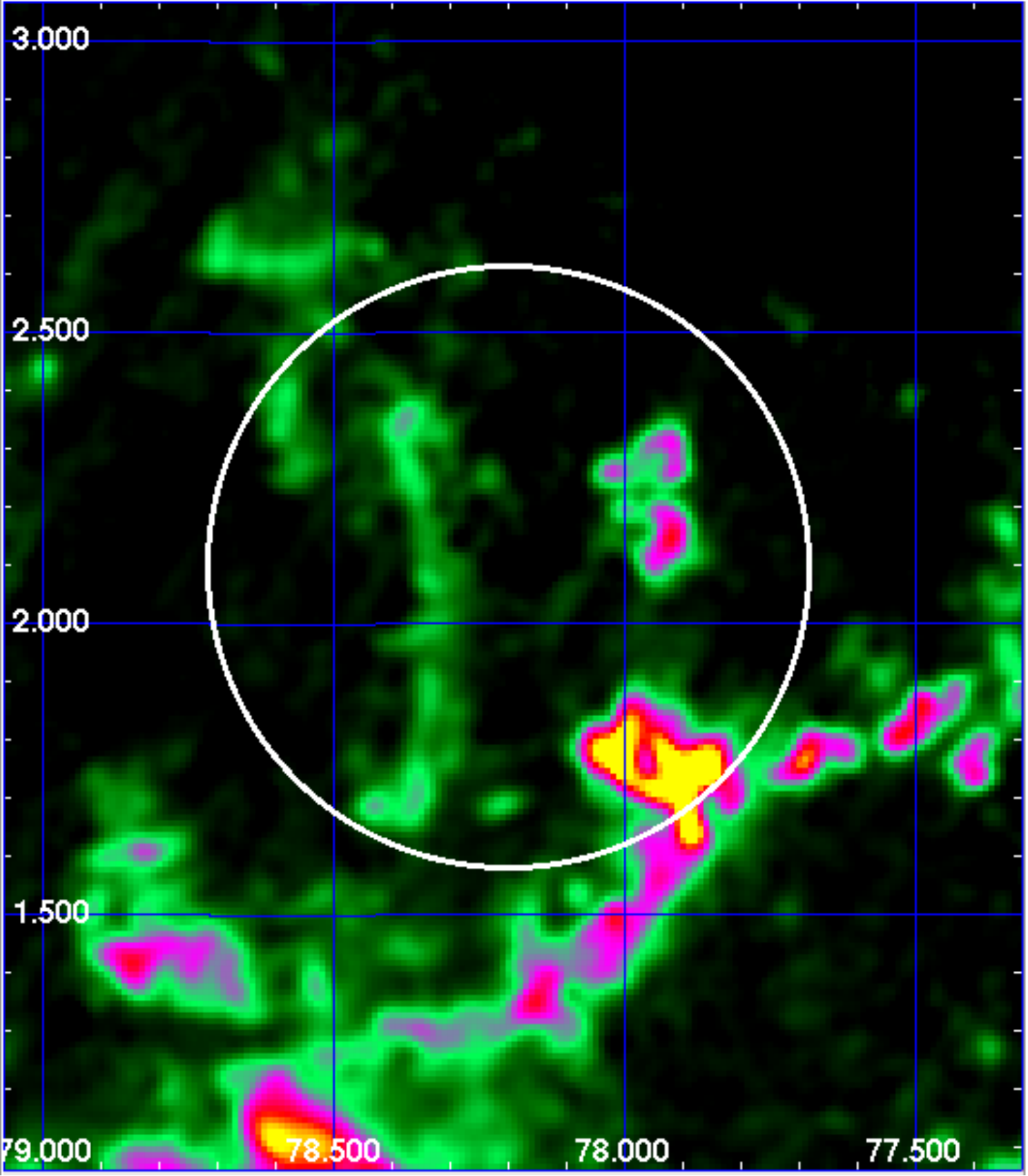} & \includegraphics[width=5.8cm]{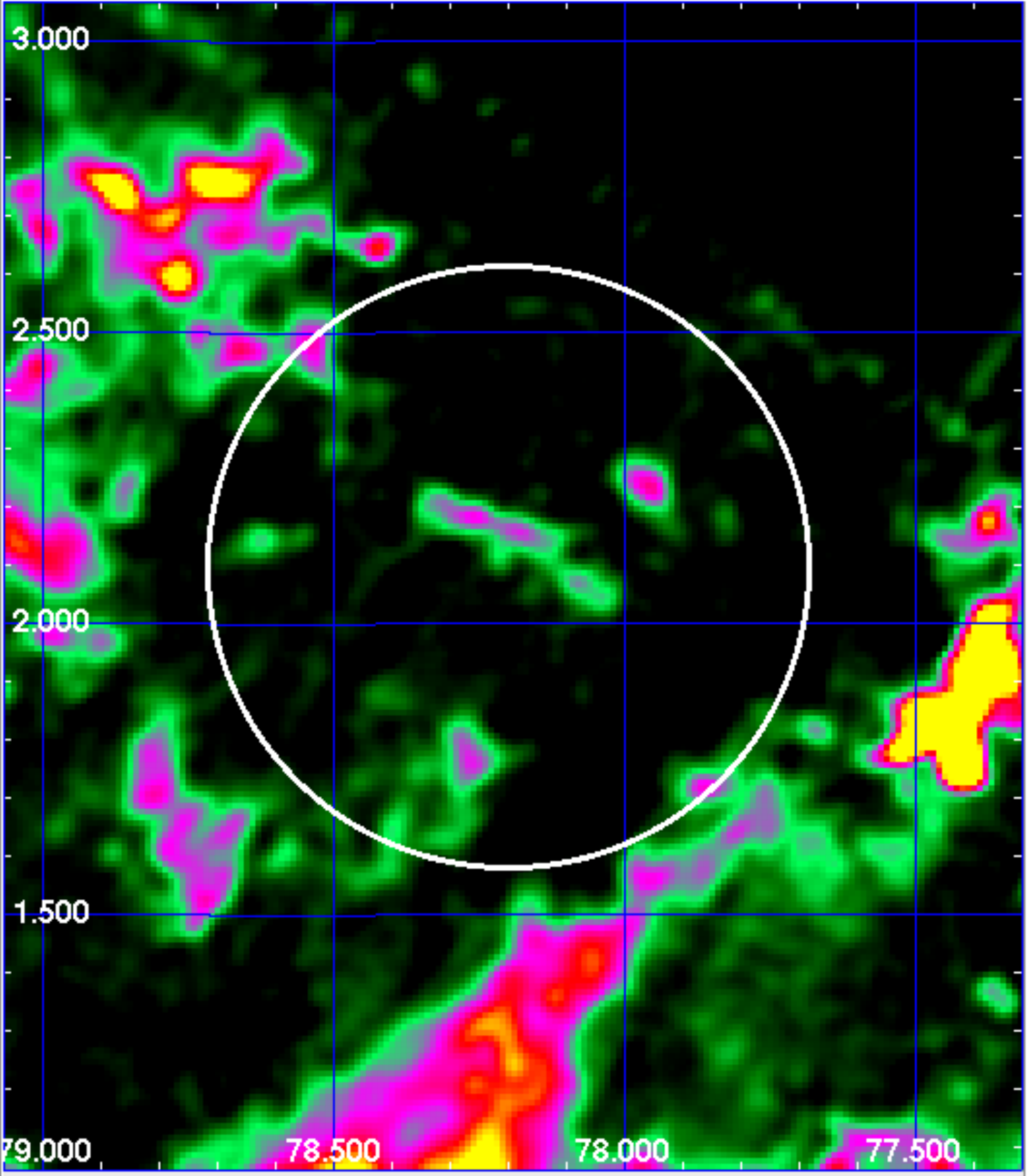} & \includegraphics[width=5.8cm]{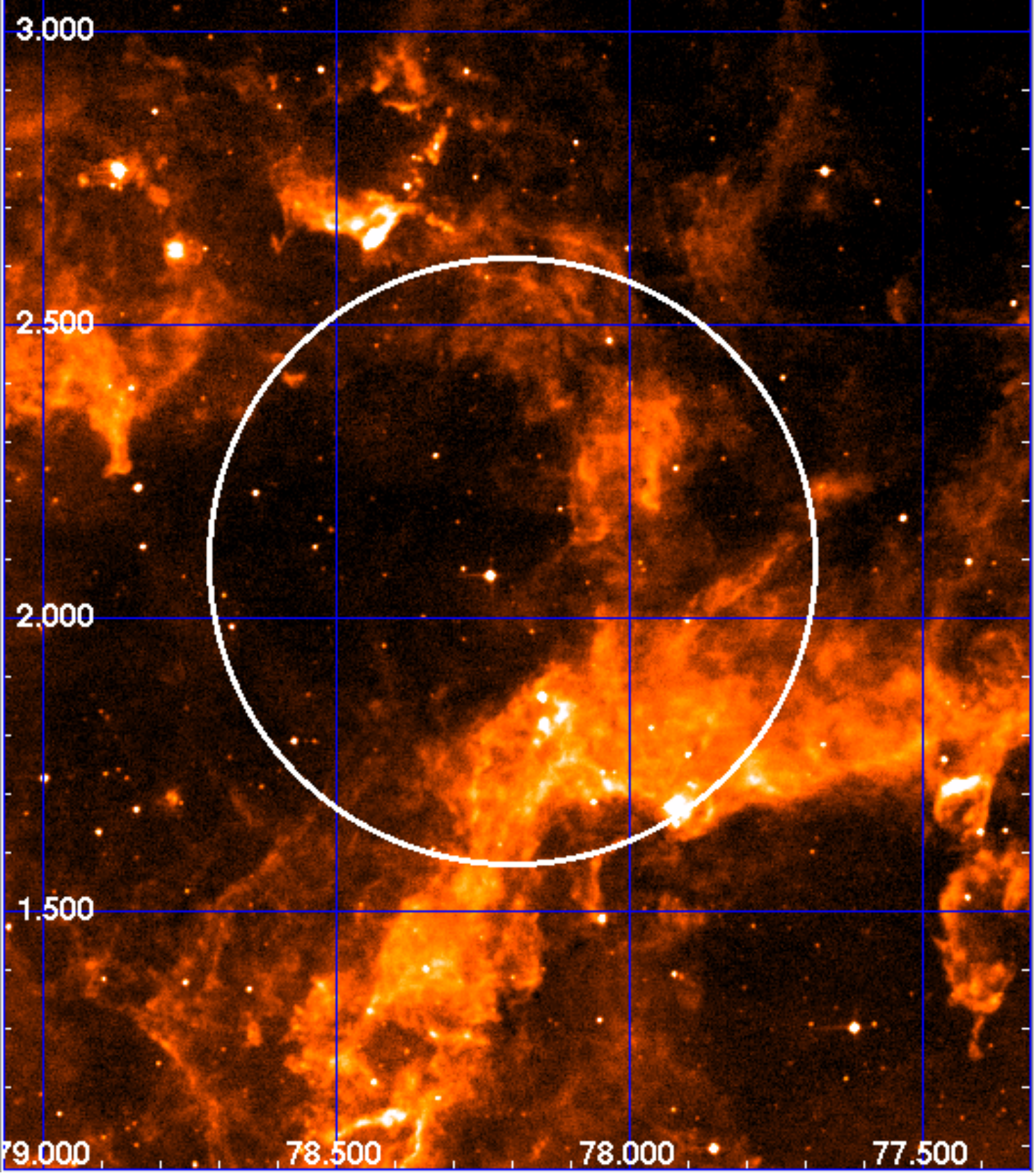}
 \end{tabular}
 \caption{\textit{Left panel}:$^{13}$CO ($J = 1\to0$) line integrated emission between -20 and 0 $\mathrm{km~s^{-1}}$ from FCRAO survey in Galactic coordinates. \textit{Central panel}:$^{13}$CO ($J = 1\to0$) line integrated emission between 0 and 20 $\mathrm{km~s^{-1}}$ from FCRAO survey in Galactic coordinates. \textit{Right panel}: MSX (band A) infrared image at 8.23 $\mu$m (pixel size = 6$''$) in Galactic coordinates. The white circle indicates the radio shell extent of the Gamma Cygni SNR.} 
 \label{gamma_cyg_co_fcrao_IR_dust}
 \end{center}
 \end{figure*}

Assuming that CO and dust are good tracers of gas, we can estimate the pattern of the gas distribution by CO line emission and IR emission from the dust. The $^{13}$CO ($J = 1\to0$) line emission from FCRAO survey, related to a positive velocity integration (0 to 20 $\mathrm{km~s^{-1}}$), shows a nearly empty region within the whole SNR extent (central panel in Fig.~\ref{gamma_cyg_co_fcrao_IR_dust}). Interestingly, the negative velocity integration map (-20 to 0 $\mathrm{km~s^{-1}}$, left panel in Fig.~\ref{gamma_cyg_co_fcrao_IR_dust}) shows a dense cloud that partially overlaps the \grid \gray diffuse emission along the SE rim of the shell.
If we refer to the gas velocity distribution in the Local Arm inferred from H~I kinematics \citep{ladouceur_08}, this cloud -- having a negative velocity range -- could represent an approaching concentration of gas located in the surrounding of the SNR.

The IR image (right panel of Fig.~\ref{gamma_cyg_co_fcrao_IR_dust}) shows the cosmic dust distribution in the SNR angular region.
This is an image in the 8.23 $\mu$m wavelength (MSX, band A) of the region; this band may contain continuum thermal emission from dust at $\sim$400 K \citep{ladouceur_08}. We can notice an arc-like structure (extending from $\sim$(78.0, 2.0) to
$\sim$(78.4, 2.7)) and a feature, located South-East of the synchrotron shell, associated by \citet{ladouceur_08} to a similar thermal feature visible at radio wavelengths (see Fig.~\ref{gamma_cyg_radio_agilecon}). It is interesting to note that this hotspot is spatially consistent also with a bright \gray excess detected by the \grid above 400 MeV. 

\subsection{High energy imaging: X-ray, HE and VHE \grays}
By analyzing  the X-ray data from ASCA, \citet{uchiyama_02} found that, at energies below 3 keV, the bulk of the X-ray flux from the SNR can be well described by thermal emission from a plasma with a temperature of $kT_e \simeq$ 0.5--0.9 keV. In addition there is an extremely hard X-ray component from several clumps located in the NW part of the SNR. They found that this unusually hard spectra can be naturally interpreted with electron non-thermal Bremsstrahlung emission.

In the same region of the shell a bright source, VER J2019+407, is visible above 10 GeV by \lat \citep{fraija_16,ackermann_17} and at TeV energies by VERITAS \citep{aliu_13,abeysekara_18}. This high-energy feature is consistent with the NW bright excess of the radio synchrotron shell.

On the other hand, AGILE detected a different morphology of the emission above 400 MeV, consistent with SE bright feature of the radio shell (DR4, see Fig. \ref{gamma_cyg_radio_agilecon}). Nevertheless, in order to have a more conservative approach, taking into account the AGILE PSF above 400 MeV, we considered in our extended analysis the average off-pulse emission detected by the \grid from the whole SNR angular extent.

%%%%%%%%%%%%%%%%%%%%%%%%%%%%%%%%%%%%%%%
\section{Multiwavelength spectrum and modeling}
%%%%%%%%%%%%%%%%%%%%%%%%%%%%%%%%%%%%%%%

For the analysis of the SNR spectral energy distribution (SED), we considered the \grid spectral data together with data from the literature, referred to the whole extent of the SNR shell. In our models we took into account only the non-thermal contributions to the overall multiwavelength SED.
Together with the AGILE data, we considered radio data (from \citealp{higgs_77a,higgs_77b,pineault_90,wendker_91,zhang_97,kothes_06,gao_11}) and \lat data (from \citealp{ackermann_17} and \citealp{abeysekara_18}). The \gray data from \lat are related to an energy band for which the emission of the pulsar is naturally turned off (cut-off energy $E_{cut-off} \simeq 2.4$ GeV). Thus, the high-energy \gray emission can be assumed to belong to the SNR.

An X-ray analysis of the region was presented by \citet{hui_15}, showing Chandra and XMM-Newton data possibly related to the SNR emission. Since no evidence of non-thermal radiation was found by the authors, we did not account for their spectra in our non-thermal modeling.

During the last few years, several works on the \gray emission from this SNR \citep{lande_12, aliu_13, fraija_16,ackermann_17,abeysekara_18} have been published but, because of the presence of the pulsar, all the analyses were focused at energies above 1 GeV, where its contribution becomes subdominant. In this paper, the off-pulsed analysis of the \gray emission from the region allowed us to extract the SNR contibution down to hundreds of MeV energies. In this section, we try to model the data with both hadronic and leptonic populations, following the re-acceleration/acceleration model described in \cite{cardillo16} and tuning several parameters of the systems according to what we expect in this kind of environment.

For the likelihood analyses of the off-pulsed extended emission, we used the whole radio emitting region as template. Consequently, to obtain the simplest model as possible, we assumed that the \gray emission detected by AGILE (this paper) and \lat \citep{ackermann_17,abeysekara_18} can be explained by an overall model, based on the same parameters.

As it is largely explained in \cite{cardillo16}, we have a lot of parameters to consider. To reduce the degrees of freedom, we have taken some parameters as fixed as found in the literature: the distance, $d=1.5$ kpc, and the age, $t_{age}=6.5\times10^{3}$ yrs  \citep{fraija_16}; the SNR dimension, $D=64'$ \citep{higgs_77a}. We also limited the shock velocity values to a range that takes into account recent estimations over the last years, $750\div1500$ km/s (\citealp{fraija_16} and references therein).

All the parameters used in our modeling are physically reasonable and tuned to get the best emission model for our source \citep[see][for a description of all the dependencies of these parameters]{cardillo16}: the interaction time (between the shock and the surrounding medium) $t_{int}$, the shock velocity $v_{sh}$, the initial magnetic field $B_{0}$, the magnetic field of the compressed downstream medium $B_m$, the initial density $n_0$, the compressed downstream density $n_m$, the momentum spectral index $\alpha$ depending on the compression ratio $r$ $\left(\alpha=\frac{3r}{r-1}\right)$, the maximum energy of accelerated particles $E_{max}$, the filling factor $f_{V}$ defined in the source volume $V=f_{V}\frac{4}{3}\pi R^{3}$, the magnetic perturbation correlation length $L_{c}$, and the acceleration efficiency $\xi_{CR}$.

The leptonic processes that we took into account are: synchrotron, Bremsstrahlung, and inverse Compton (IC) processes (both on cosmic microwave background, CMB, and interstellar radiation, ISR). The hadronic emission model estimates the \gray emission from the decays of the $\pi^0$ mesons ($\pi^0 \to \gamma \gamma$) produced in proton-proton inelastic scatterings, and from secondary electrons produced in charged pion decays.

In our modeling of the SNR SED, we considered two main scenarios. In the first one, we take into account the emission from the whole SNR angular extent: in the \gray band we consider the AGILE spectrum together with the \lat data from \citet{ackermann_17} and \cite{abeysekara_18}, both calculated over the $\sim$1 degree extent of the SNR. We present both a single population and a double population emission models. In the second scenario, we focus on the NW rim of the shell, at the position of the VER J2019+407 source: in the \gray band we consider the \lat spectrum from \citet{fraija_16} and \citet{abeysekara_18}, and the VERITAS spectrum from \citet{aliu_13} and \citet{abeysekara_18}.

\subsection{The overall model}
We know that in our Galaxy there is a sea of Galactic Cosmic Rays (CRs) that could be re-accelerated by first order Fermi mechanism in correspondence of the SNR shocks \citep[][and references therein]{cardillo16}. Consequently, we tried to compute radio and \gray emission due to re-acceleration of pre-existing CRs. However, the initial density value is too low to provide a re-acceleration contribution sufficiently high to explain the AGILE \gray detection.

Thus, we neglected re-acceleration and we focused our analysis on possible freshly accelerated particles, analyzing both hadronic and leptonic dominant emission.

As described above, the dataset refers to the whole SNR extent. In the SED modeling, we also plotted the ASCA and VERITAS data in gray as reference (and upper limit) values, even if they are related to the NW rim of the shell.

%%%%%%%%%%%%%%%%%%%%%%%%%%%%%%%%%%%%%%%
\subsubsection{Hadrons}
%%%%%%%%%%%%%%%%%%%%%%%%%%%%%%%%%%%%%%%
%---------------------------------------Hadronic
\begin{figure*}[!h]
 \begin{center}
 \begin{tabular}{cc}
 \includegraphics[width=8.5 cm]{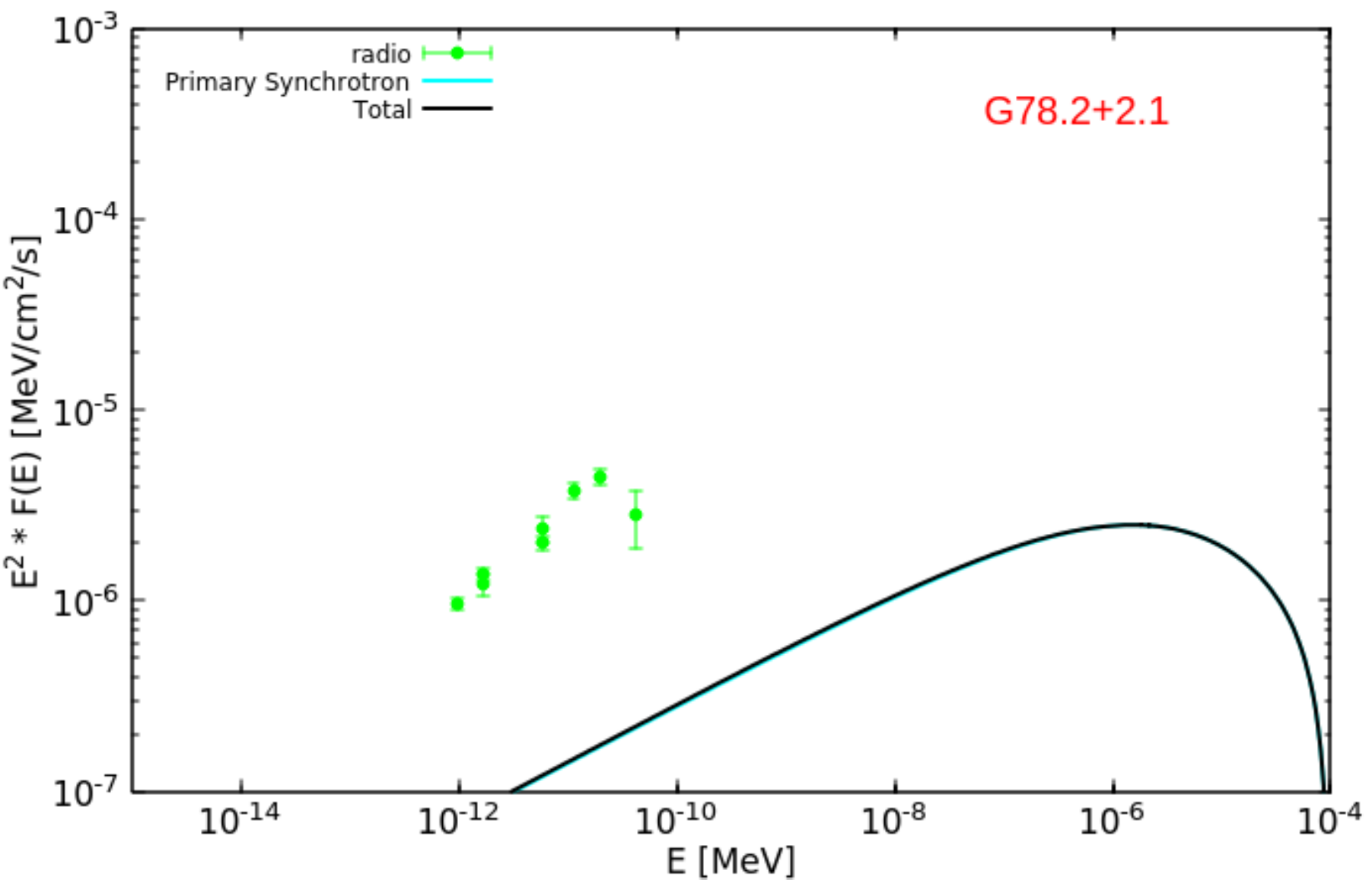} & 
 \includegraphics[width=8.5 cm]{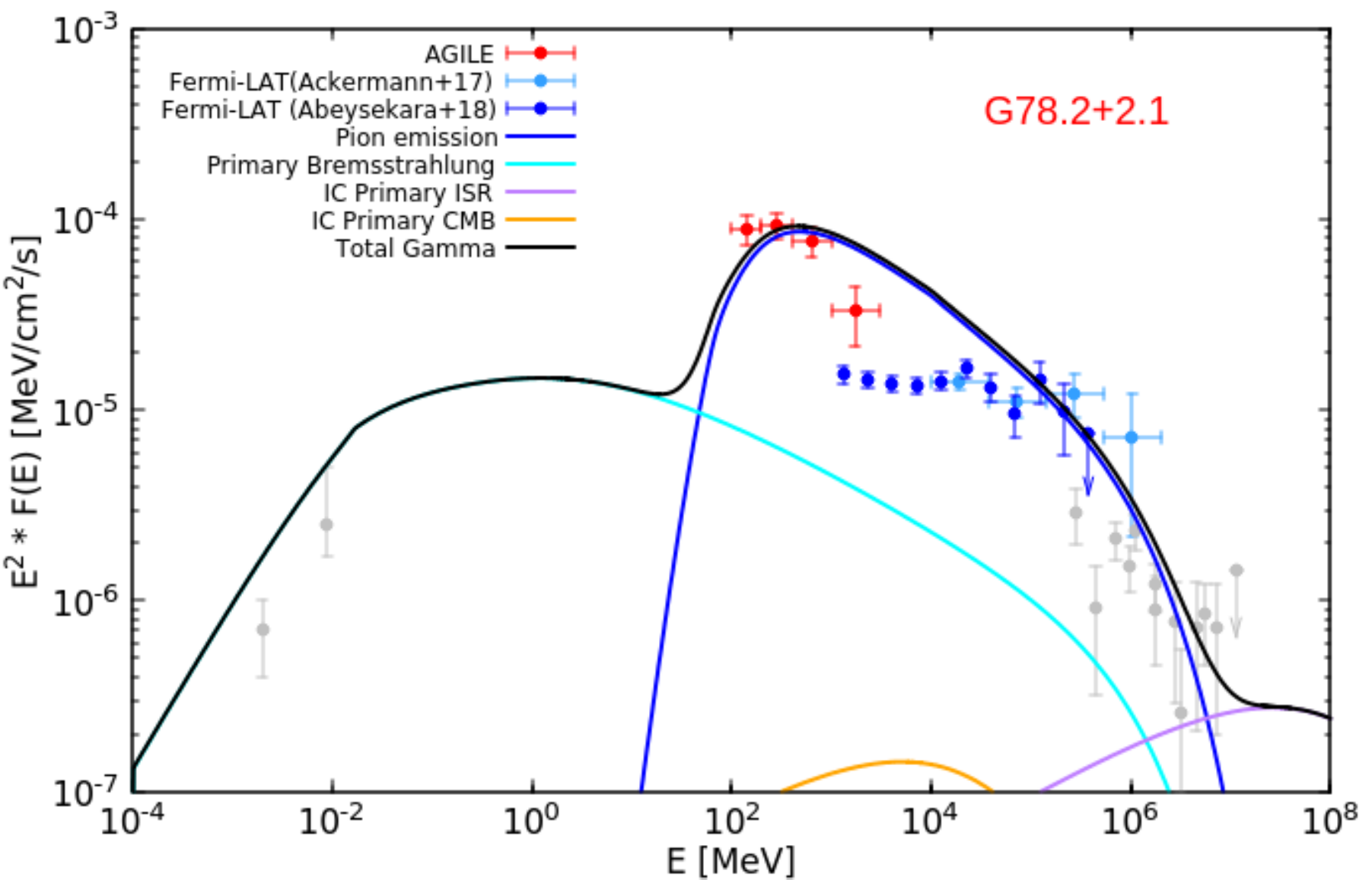}
 \end{tabular}
 \caption{Single population hadronic model. \textit{Left panel}: Radio spectrum, average emission from the whole extent of the SNR G78.2+2.1 \citep{higgs_77a,higgs_77b,pineault_90,wendker_91,zhang_97,kothes_06,gao_11} with the corresponding leptonic component of a hadron-dominated emission model. \textit{Right panel}: \grid (100 MeV -- 3 GeV) and \lat (1--500 GeV from \citealp{abeysekara_18}, 0.01--2.00 TeV from \citealp{ackermann_17}) spectra, average emission from the whole extent of the SNR with the corresponding hadron-dominated emission model. Grey data points in background are related to the emission from the region of VER J2019+407 only.} 
 \label{Fig:Model_Hadro}
 \end{center}
 \end{figure*}
%-------------------------------------------------------------------
In Fig.~\ref{Fig:Model_Hadro} we show both radio (left) and \gray (right) SED with our best hadronic acceleration emission model obtained with $t_{int}=0.9\times t_{age}$, $v_{sh} = 850$ km/s, $L_{c} = 0.2$ pc, $B_{0}=8.5$ $\mu$G, $B_{m}=21.9$ $\mu$G, $n_{0}=10$ cm$^{-3}$, and $n_m=31.4$ cm$^{-3}$. We used a simple power-law distribution for both nuclei and electrons, with a momentum index $\alpha=4.4$. The derived cut-off particle energy is at $E_{max}=5.9$ TeV/n ($\sim$ 590 GeV for \gray). With low CR acceleration efficiency $\xi_{CR}=0.2\%$, the \gray emission from $\pi^{0}$ decay can explain both AGILE and \lat \citep{ackermann_17,abeysekara_18} data points. The Bremsstrahlung contribution becomes dominant at $E< 10$ MeV and IC emission is completely negligible since the high density value.

However, this model cannot in any way explain the detected radio emission, even because only primary electrons contribute to radio spectrum, since the density value is not sufficient to give a higher secondary electron production from $\pi^{\pm}$-decays. The expected Bremsstrahlung emission slightly exceeds the ASCA data. Furthermore, the modeling of the \gray spectrum is not accurate basing exclusively on this simple hadronic scenario.

 %%%%%%%%%%%%%%%%%%%%%%%%%%%%%%%%%%%%%%%
\subsubsection{Leptons}
%%%%%%%%%%%%%%%%%%%%%%%%%%%%%%%%%%%%%%%
\subsubsection*{Single population}

\begin{figure*}[!hb]
 \begin{center}
 \begin{tabular}{cc}
 \includegraphics[width=8.5 cm]{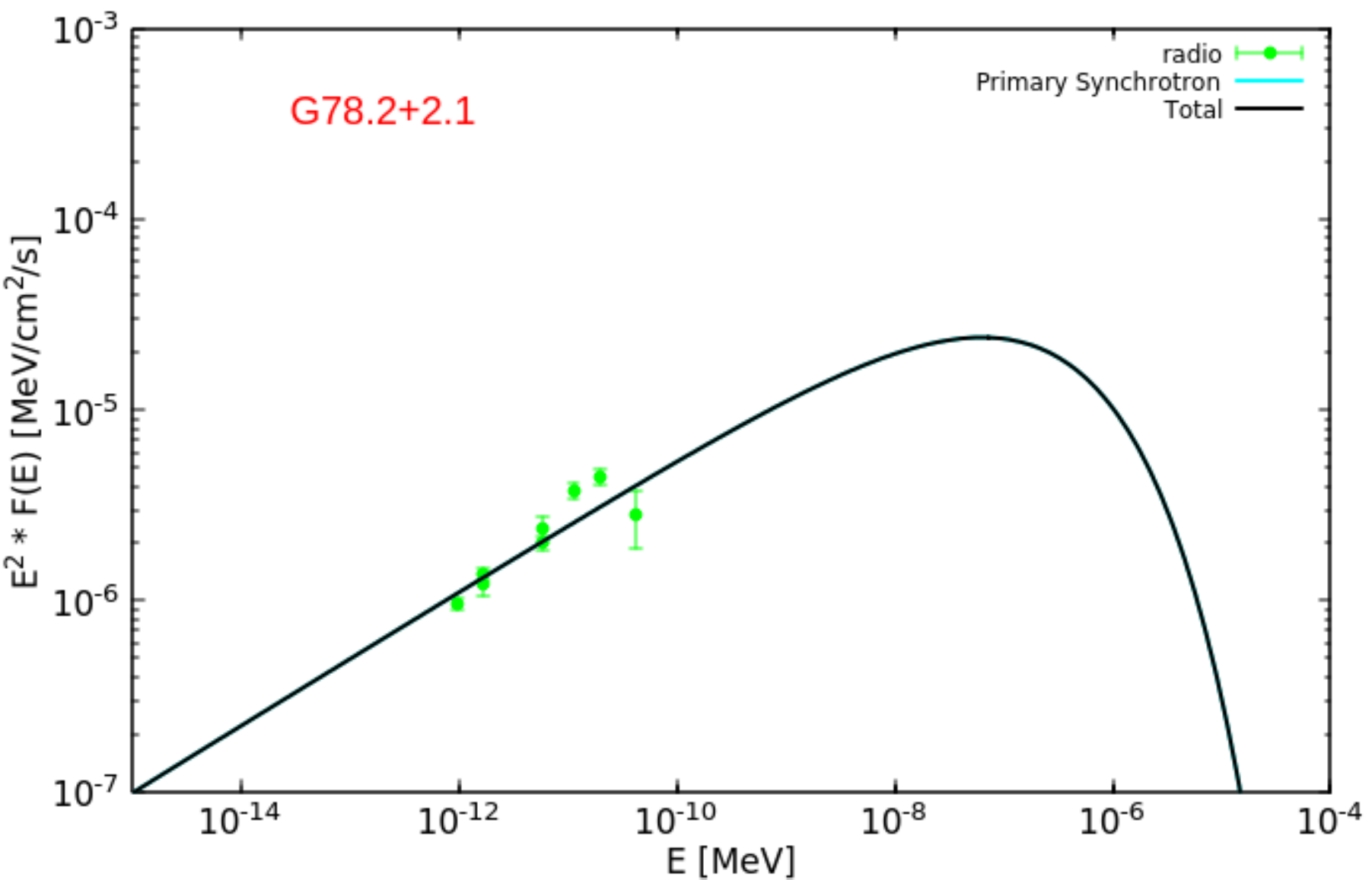} & 
 \includegraphics[width=8.5 cm]{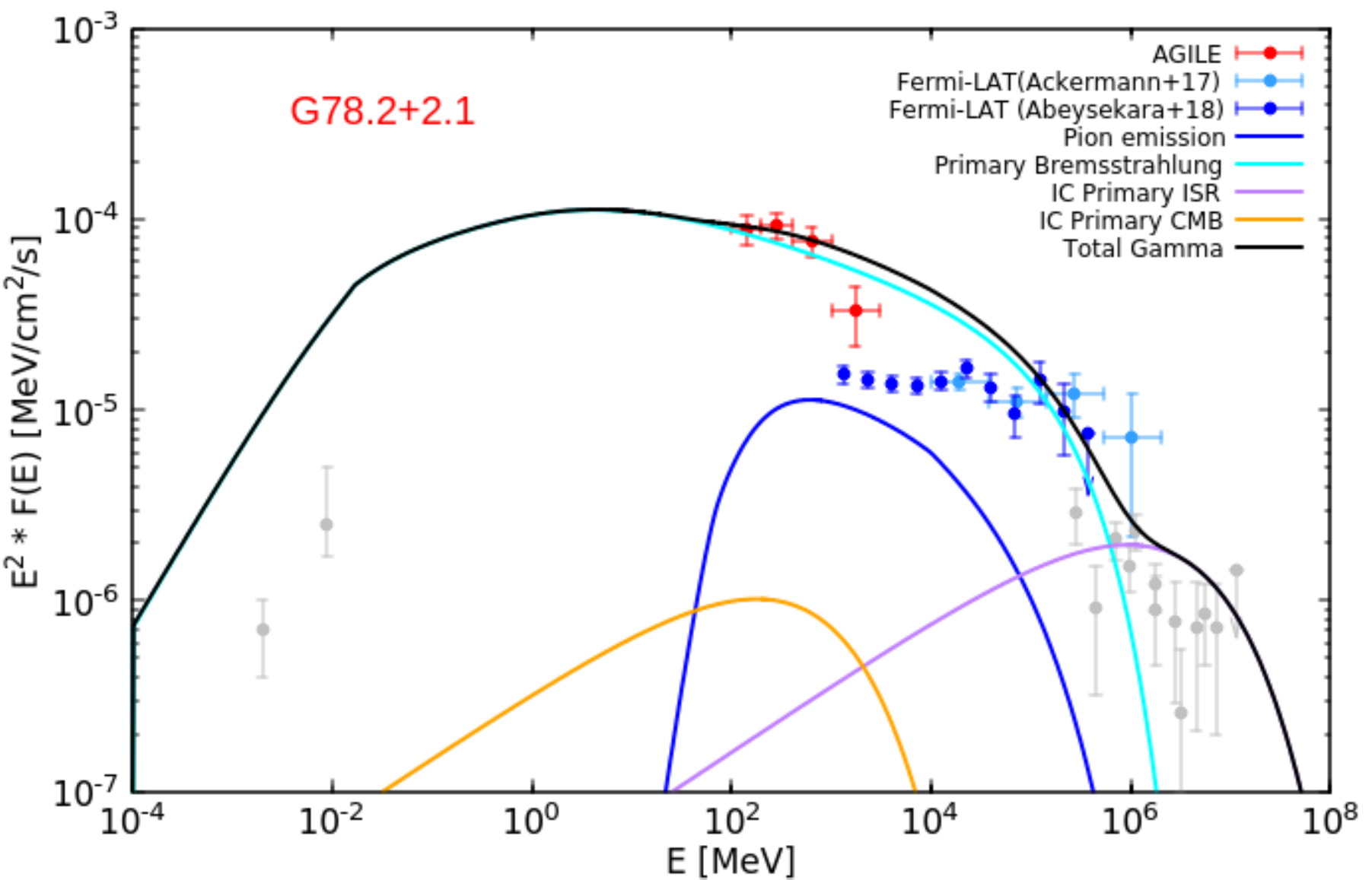}
 \end{tabular}
 \caption{Single population leptonic model. \textit{Left panel}: Radio spectrum, average emission from the whole extent of the SNR G78.2+2.1 (same as in Fig.~\ref{Fig:Model_Hadro}) with the corresponding leptonic component of a lepton-dominated emission model. \textit{Right panel}: \grid and \lat spectra (the same as in Fig. \ref{Fig:Model_Hadro}) with the corresponding lepton-dominated emission model.} 
 \label{Fig:Model_Lepto_single}
 \end{center}
 \end{figure*}

In order to fit both \gray and radio emission, we consider a leptonic emission model, shown in Fig.~\ref{Fig:Model_Lepto_single}, with an electron/proton ratio equal to 1. 

This model was obtained with parameters slightly different from the hadronic one, but always consistent with the physics of the system; $t_{int}=0.8\times t_{age}$, $v_{sh} = 850$ km/s, $L_{c} = 0.8$ pc, $B_{0}=9.5$ $\mu$G, , $B_{m}=25.6$ $\mu$G, $n_{0}=10$ cm$^{-3}$, and $n_m=33.1$ cm$^{-3}$. The simple power-law distribution has a harder momentum index, $\alpha=4.3$, implying also a lower efficiency value, $\xi_{CR}=0.02\%$. The derived particle maximum energy is  $E_{max} = 5.2$ TeV/n ($\sim$ 520 GeV for \grays).

As shown in Fig.\ref{Fig:Model_Lepto_single}, the leptonic model can explain both radio and \gray emission, even though the GeV domain is not accurately fitted. We note that the \gray emission is dominated by the Bremsstrahlung emission, strongly related to the density value, and the IC contribution is higher (than the hadronic scenario) because of the higher electron/proton ratio. However, the expected emission in the X-ray band is much higher than the ASCA data, indicating that this model is not accurate to describe the high-energy non-thermal radiation from the SNR.

%--------------------------------------- Leptonic single
 %---------------------------------------

\subsubsection*{Double population}
We found that both hadronic and leptonic models with a single population cannot accurately explain the radio, X-ray and \gray emission from \gcyg. Thus we considered the possibility that the MeV-GeV \gray emission detected by AGILE is due to a population interacting with a higher density medium, and the GeV-TeV emission detected by \lat \citep{ackermann_17,abeysekara_18} to a population interacting with a lower density medium.

Since we do not have any information about the extent of higher and lower density region of the remnant, we use the simplest approach and consider equal filling factors, $f=50 \%$ of the volume.

We consider only leptonic model for both populations, because there is no way to properly fit the radio emission with a double-population hadronic model.

\begin{figure*}[!h]
 \begin{center}
 \begin{tabular}{cc}
 \includegraphics[width=8.5 cm]{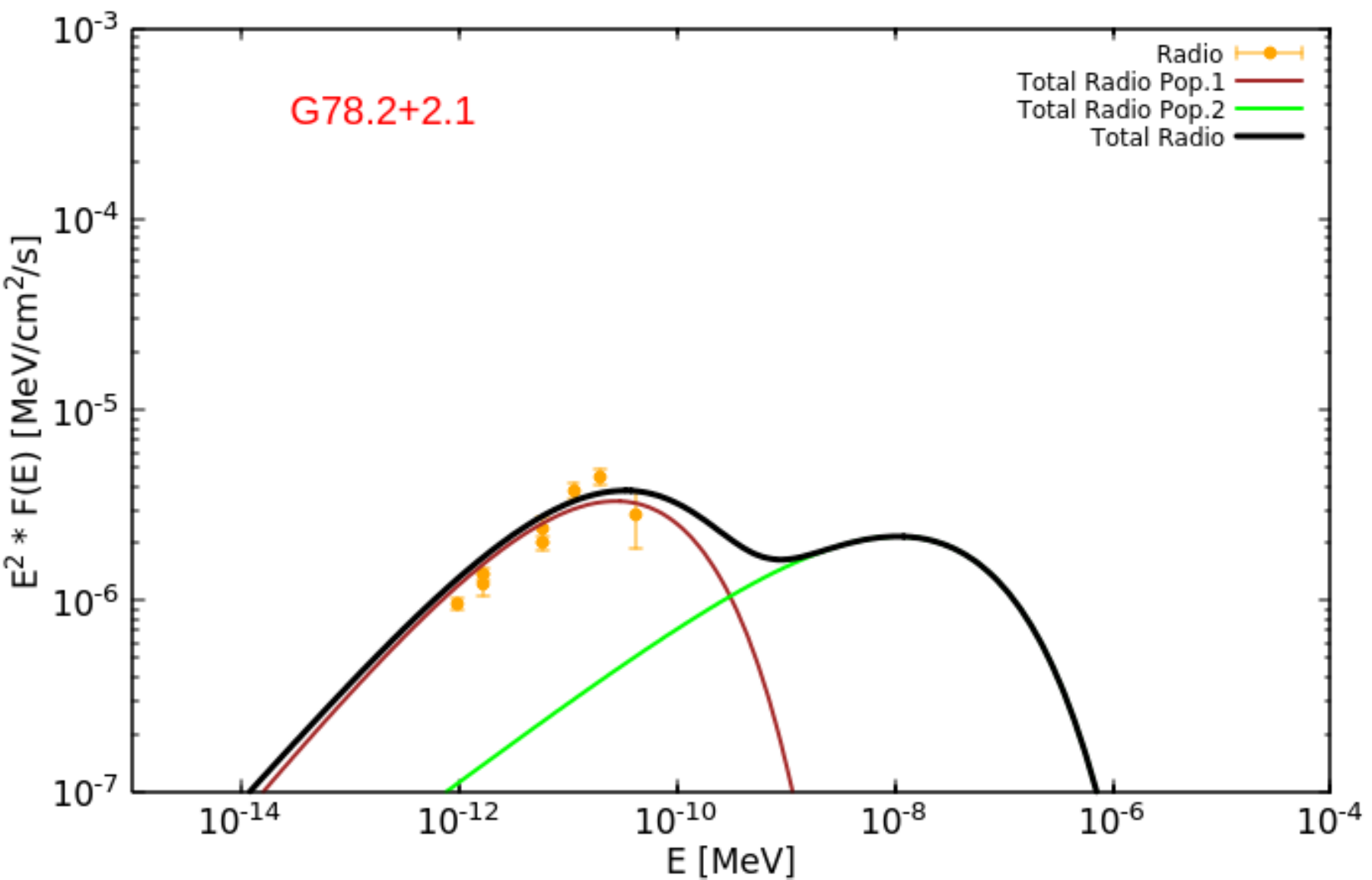} & \includegraphics[width=8.5 cm]{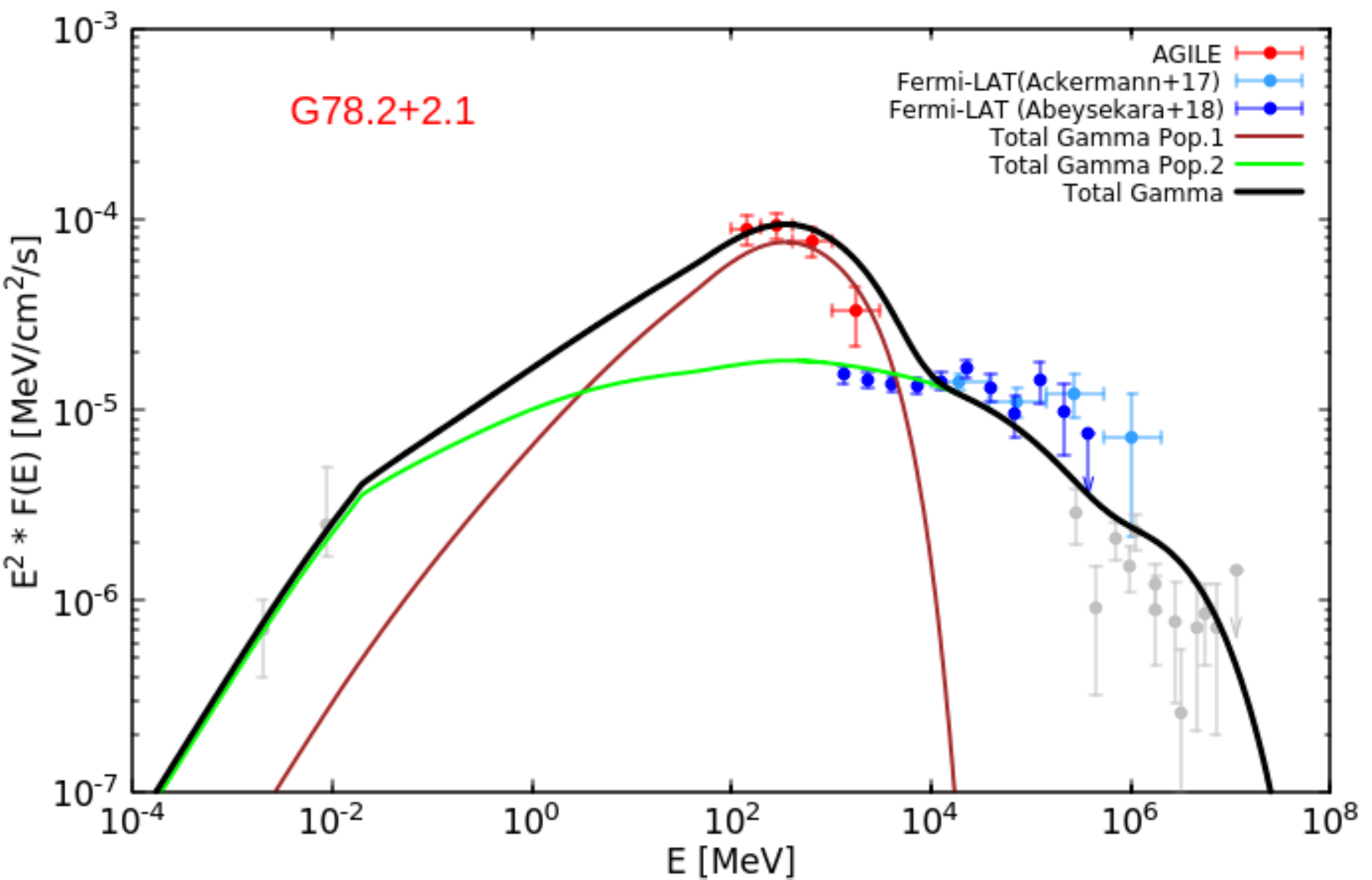}\\
 \includegraphics[width=8.5 cm]{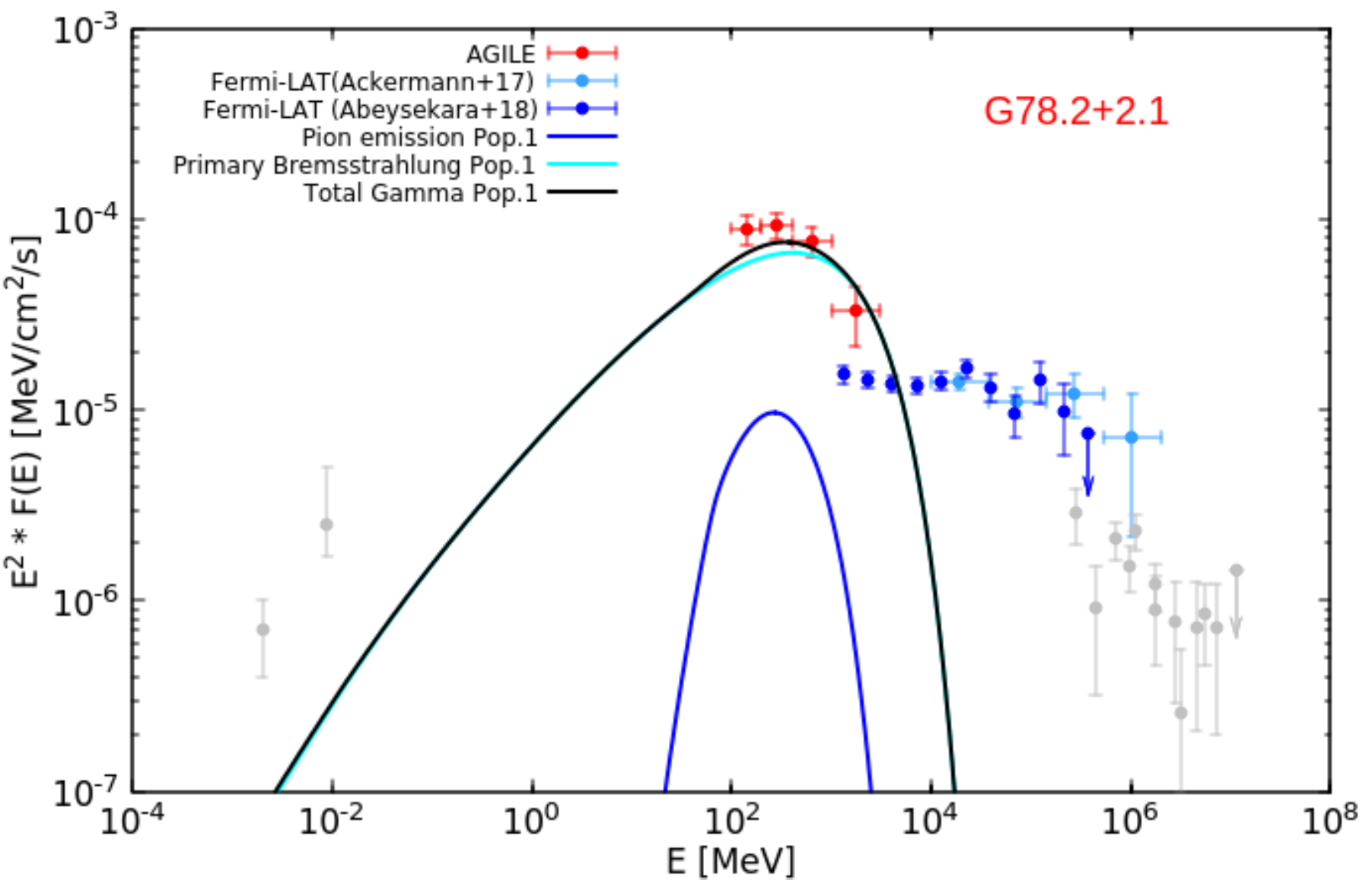} & \includegraphics[width=8.5 cm]{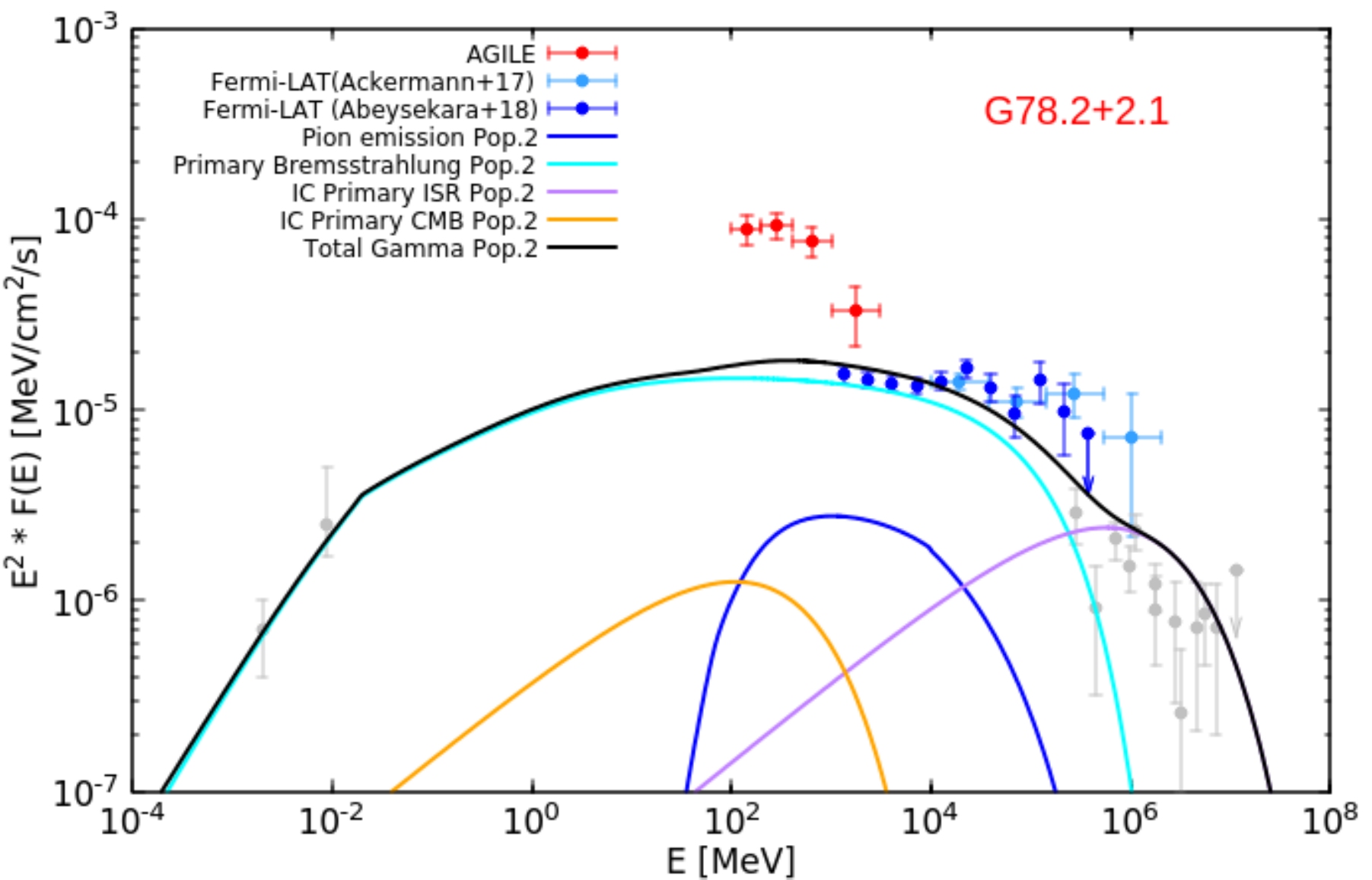}
 \end{tabular}
 \caption{\textit{Top panels}: Double population leptonic model. \textit{Top-left panel}: Radio spectrum, total emission in black from the two leptonic populations, Pop. 1 (brown) and Pop. 2 (green) \textit{Top-right panel}: \grid (100 MeV -- 3 GeV), \lat (whole SNR, 1--500 GeV from \citealp{abeysekara_18}, 0.01--2.00 TeV from \citealp{ackermann_17}), with the corresponding total \gray emission from two leptonic populations, Pop. 1 (brown) and Pop. 2 (green). \textit{Bottom panels}: High-energy emission model components in detail for the two populations. \textit{Bottom-left panel}:  first population. \textit{Bottom-right panel}: second population.}
 \label{Fig:Model_Lepto_double}
 \end{center}
 \end{figure*}

In Fig.\ref{Fig:Model_Lepto_double}, our best double-population model is shown. We note how the first population (high density) can explain the AGILE and radio datasets; from Fig.~\ref{gamma_cyg_radio_agilecon}, it is evident that the most prominent radio peak is strongly correlated with the AGILE \gray detected emission. On the other hand, the second population (low density) properly fits the \lat emission above 1 GeV and the X-ray emission detected by ASCA \citep{uchiyama_02}.

For the first population, we assumed $t_{int}=0.3\times t_{age}$, $n_{0}=50$ cm$^{-3}$, $v_{sh} = 750$ km/s, $L_{c} = 2.9$ pc, $B_{0}=21$ $\mu$G, $\alpha = 3.6$ and $\xi_{CR}=1.4\times10^{-5}$, obtaining $B_{m}=104$ $\mu$G, $n_m=300$ cm$^{-3}$, $E_{max}= 2 $ GeV/n.

For the second population, we assumed $t_{int}=t_{age}$, $n_{0}=2.5$ cm$^{-3}$, $v_{sh} = 920$ km/s,  $L_{c} = 1.0$ pc, $B_{0}=2.4$ $\mu$G,  $\alpha = 4.2$, and $\xi_{CR}=1.1\times10^{-1}$, obtaining $B_{m}=7$ $\mu$G,  $n_m=9$ cm$^{-3}$ and $E_{max}= 363 $ GeV/n.

%--------------------------------------- Leptonic double
 %---------------------------------------

\subsection{Modeling the VER J2019+407 source}
\begin{figure*}[!ht]
 \begin{center}
 \begin{tabular}{cc}
 \includegraphics[width=8.5 cm]{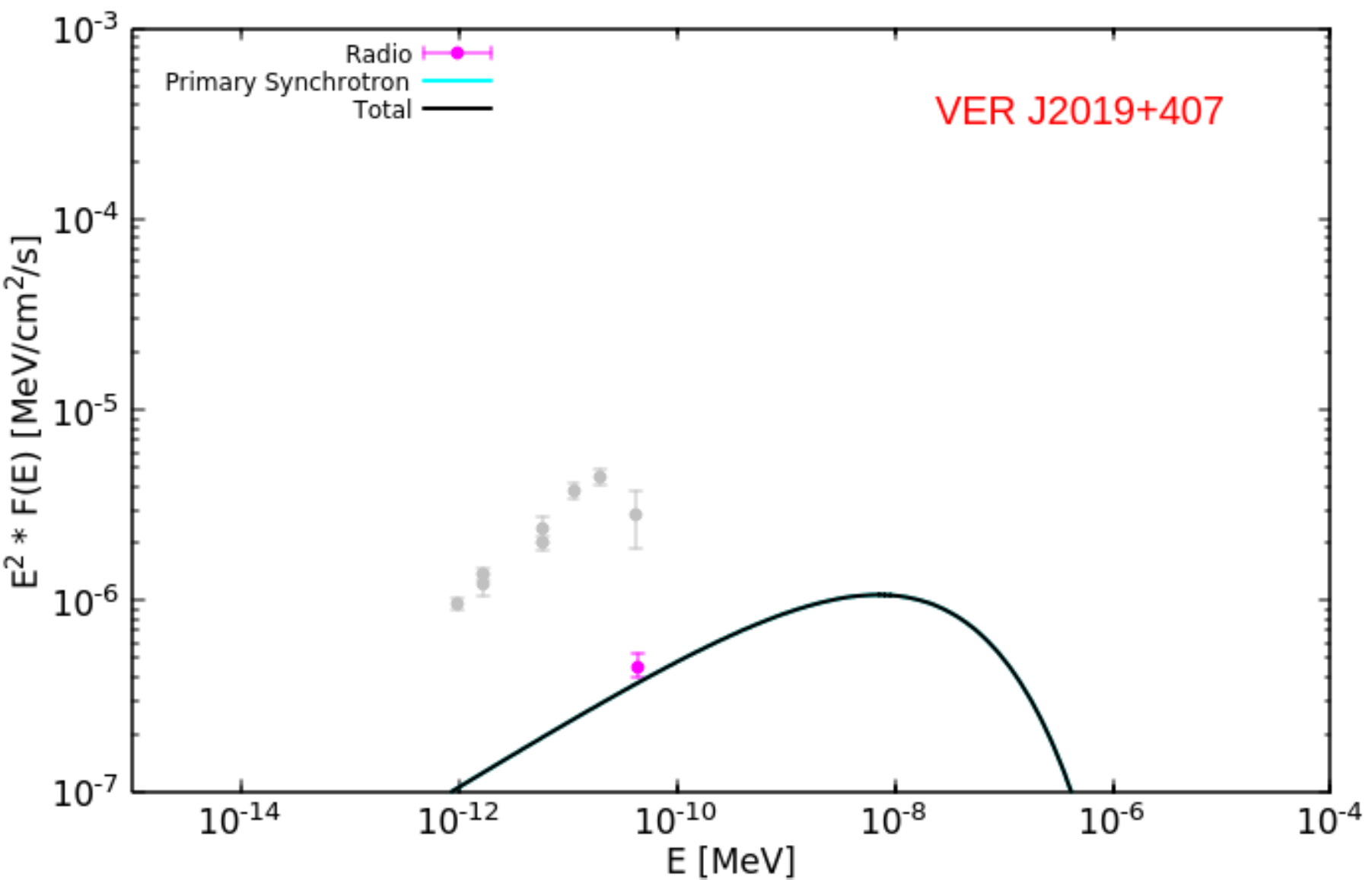} & 
 \includegraphics[width=8.5 cm]{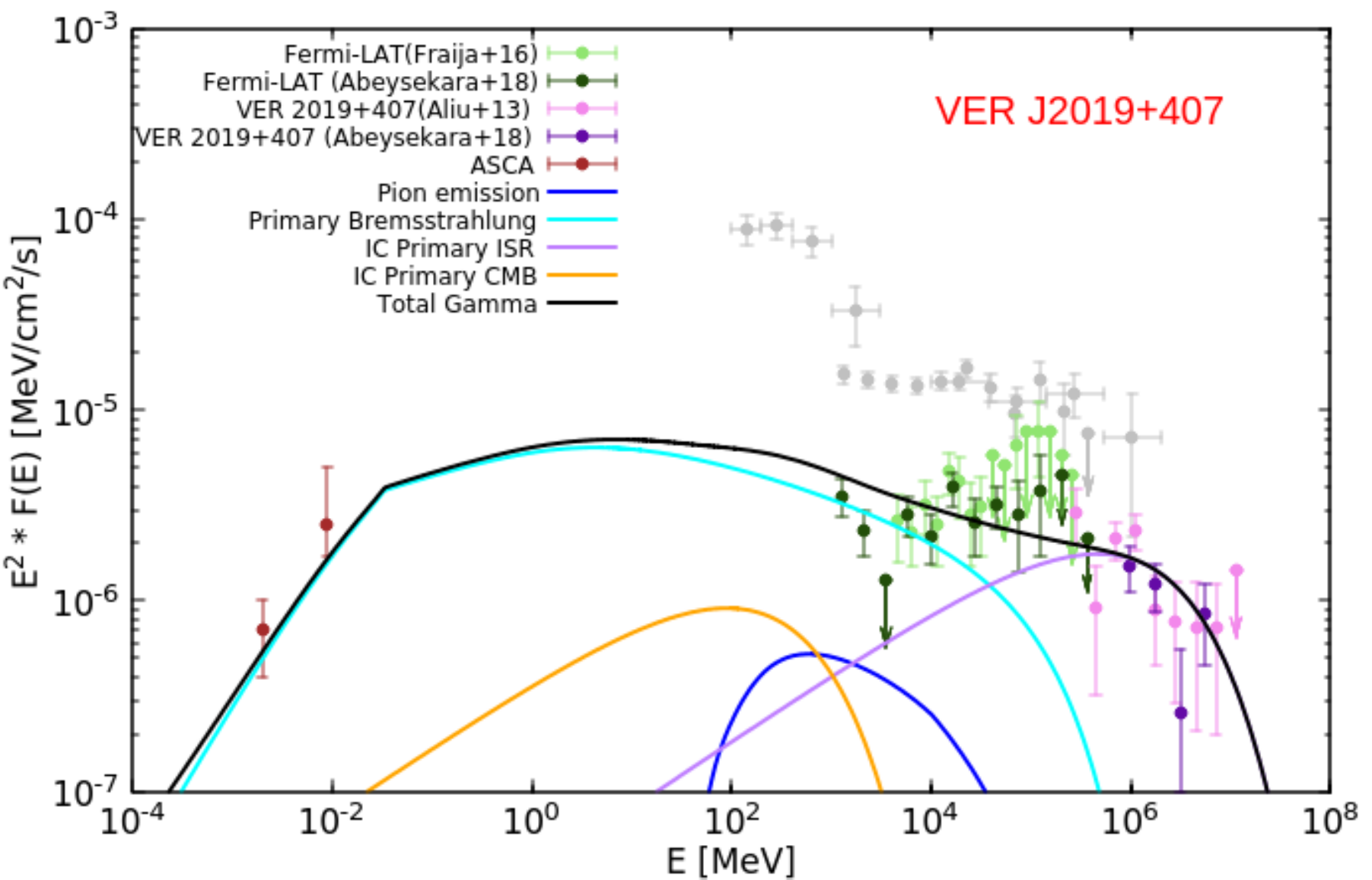}
 \end{tabular}
 \caption{Leptonic model for the $\gamma$-ray emission detected in the North-Western side of the SNR (VER J2019+407, see Fig.~\ref{gamma_cyg_radio_TeV}). \textit{Left panel}:  Radio emission from the GeV-TeV region \citep{higgs_77a} with the corresponding leptonic component of a lepton-dominated emission spectrum. \textit{Right panel}: ASCA (2-10 keV, hard X-ray clumps, \citealp{uchiyama_02}, \lat (1--500 GeV from \citealp{abeysekara_18}, 4--300 GeV from \citealp{fraija_16}) and VERITAS \citep{aliu_13,abeysekara_18} data with the corresponding lepton-dominated emission model. Grey data points in background are related to the emission from the whole region of the SNR G78.2+2.1.}
 \label{Fig:Model_Lepto_FV}
 \end{center}
 \end{figure*}

For completeness, we also studied a model for the region associated with the VERITAS source VER J2019+407 \citep{aliu_13} in the NW rim of the SNR, considering also the corresponding \lat spectrum \citep{fraija_16,abeysekara_18}. For this model, we also took into account the X-ray spectra detected by ASCA \citep{uchiyama_02} in the NW region (hard clumps), which are characterized by a non-thermal power-law tail and positionally consistent with the VERITAS source. The radio emission in the region of the GeV-TeV source \citep{higgs_77a} was also considered in this model.\\
In order to model the emission of the GeV-TeV source VER J2019+407, we fixed a lower initial density value at $n_{0}=0.5$ cm$^{-3}$ ($n_{m}=1.7$ cm$^{-3}$), lower than the one used in the work of \cite{fraija_16} but similar to the one used in \cite{aliu_13}. Our assumption is that the remnant is interacting with this lower density region for a longer time, $t_{int}=t_{age}$, and with a higher velocity, $v_{sh}=1200$ km/s consistently with a propagation in a lower density medium. Magnetic field is lower, $B_{0}=2.1$ $\mu$G ($B_{m}=5.7$ $\mu$G), and the correlation length is higher, $L_{c}=1.9$ pc. We used a small filling factor, $f_V=20\%$, based on the angular extent of the observed GeV-TeV source.  The maximum energy is $E_{M}=6.5$ TeV/n, the momentum index is $\alpha = 4.3$, and the acceleration efficiency is $\xi_{CR}=1.0\%$. In this case, due to the lower density and the lower magnetic field value, the IC on the ISR is dominant at higher energies and fits the \gray data points detected by \lat and VERITAS (see Fig.~\ref{Fig:Model_Lepto_FV}, right panel).  
 
 %%%%%%%%%%%%%%%%%%%%%%%%%%%%%%%%% Leptonic FV
 %%%%%%%%%%%%%%%%%%%%%%%%%%%%%%%%%%
 
The Bremsstrahlung contribution is dominant at GeV and X-ray energies, well fitting the ASCA data points. Moreover, the provided radio synchrotron emission explains the radio flux within the VERITAS source \citep[][and reference within]{fraija_16}.

%%%%%%%%%%%%%%%%%%%%%%%%%%%%%%%%%%%%%%%
\section{Discussion}
\label{Sec:discussion}
%%%%%%%%%%%%%%%%%%%%%%%%%%%%%%%%%%%%%%%

Our analysis was carried out considering the whole SNR as emitter, by analyzing the data detected at radio wavelengths and \gray frequencies by AGILE and \lat \citep{ackermann_17,abeysekara_18}. The parameters used, consistent with the literature and derived according to the equations in \cite{cardillo16}, provide a negligible contribution from pre-existing re-accelerated CRs. Consequently, we considered hadronic and leptonic acceleration as the only contribution to \gray and radio emission from the SNR G78.2+2.1.

\begin{figure}[!h]
 \begin{center}
  \includegraphics[width=8.5cm]{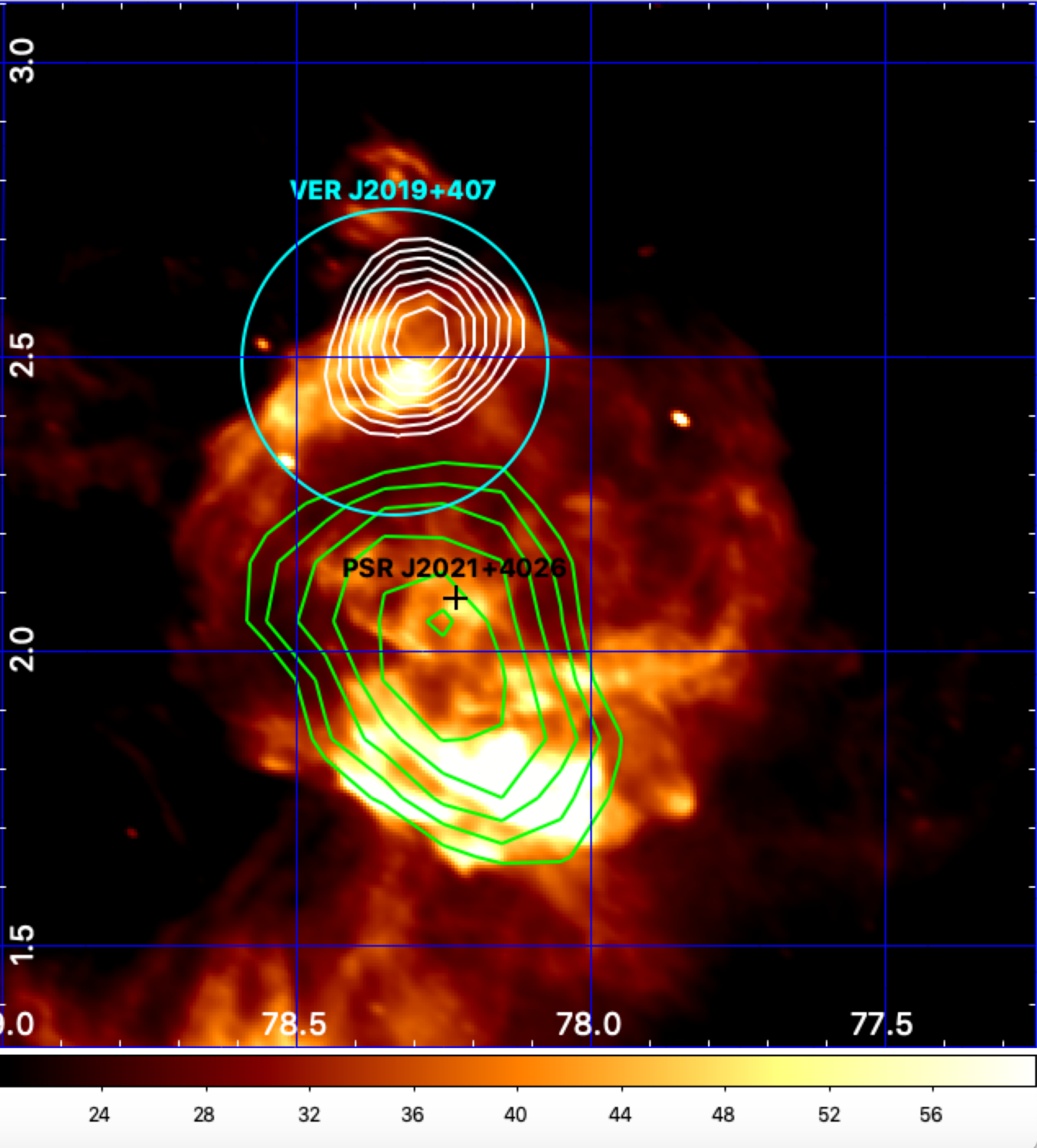}
 \caption{Radio image of the SNR G78.2+2.1 at 21.1 cm wavelength, (frequency = 1420 MHz, bandwidth = 30 MHz), pixel size = 20$''$, DRAO Radio Telescope. \textit{Green contours}: \grid contour levels, related to the off-pulse intensity map above 400 MeV. \textit{White contours}: \lat contour levels related to the background subtracted $TS$ map (photon energies greater than 10 GeV, $TS > 36$ with 3-pixel Gaussian smoothing), from \citet{ackermann_17}. The \textit{cyan circle} marks the position of the extended TeV source, VER J2019+407, as detected by VERITAS \citep{aliu_13}. The \textit{black cross} marks the position of the pulsar \gcygpsr.}
 \label{gamma_cyg_agile_fermi}
 \end{center}
 \end{figure}

We found that a single population emission model cannot properly account for the radio, X-ray and \gray non-thermal emission, whereas a double population leptonic model fits the data in a satisfactory way (see Fig.~\ref{Fig:Model_Lepto_double}). The population of accelerated leptons interacting with a high density medium can account for the MeV-GeV emission detected by AGILE (bottom-left panel of Fig.~\ref{Fig:Model_Lepto_double}), and is related to dominant Bremsstrahlung processes. On the other hand, the second population, interacting with a lower density medium, is able to account for the GeV-TeV emission detected by \lat \citep{ackermann_17,abeysekara_18}. In this case, Bremsstrahlung processes (in a lower density medium) are dominant below $\sim$250 GeV, and IC processes represent the main emission mechanism at the highest energies. We remark that, even though both the AGILE and \lat spectra are referred to the whole SNR angular extent, the first population is possibly associated with the activity of the SE rim of the shell, where most of the AGILE emission is detected; on the other hand, the second population is most likely associated with the NW rim, where most of the \lat emission is observed (see Fig.~\ref{gamma_cyg_agile_fermi}). Interesting, CO and IR emission maps (see Fig.~\ref{gamma_cyg_co_fcrao_IR_dust}) indicate a concentration of dense gas in the SE rim of the shell, supporting our hypothesis of a high-density medium in coincidence with the emission detected by AGILE.

The resulting scenario is a non-symmetrical shell where two populations of accelerated particles interact with different environments located in two regions coincident with two bright radio features: one is related with the MeV-GeV domain (SE), the other with the GeV-TeV emission (NW). According to the analysis carried out by \citet{ladouceur_08}, the SE and NW regions represent the brightest non-thermal region of the shell, indicating sites of intense synchrotron emission from shock-accelerated electrons propagating in a highly inhomogeneous medium.

We also presented the results from a further theoretical model that can account for the local emission of the \lat/VERITAS \gray source VER J2019+407. Even in this case, the re-acceleration is totally negligible. According to our study, a simple leptonic model is able to fit the data in the radio, X-ray and \gray band, by assuming a low initial density $n_0=0.5$ cm$^{-3}$ and a higher shock velocity $v_{sh}=1200$ km/s.
Our theoretical expectations for the emission model in this region are fully consistent with the X-ray ``hard clump'' reported by \citet{uchiyama_02} in the NW rim of the shell: non-thermal Bremsstrahlung in a lower density medium (with respect to the SE rim) is consistent with the hard-power law tail detected by ASCA at 2--10 keV. As in \citet{fraija_16}, our model requires a lower density medium with respect to the 10 cm$^{-3}$ adopted by \citet{uchiyama_02}, which would overpredict the GeV-TeV emission in this region. Nevertheless, in order to properly fit the data, we assumed a higher shock velocity with respect to the one derived by the authors of that paper.

%%%%%%%%%%%%%%%%%%%%%%%%%%%%%%%%%%%%%%%
\section{Conclusions}
%%%%%%%%%%%%%%%%%%%%%%%%%%%%%%%%%%%%%%%

In this paper we analyzed the \gray emission in direction of the SNR G78.2+2.1, by selecting only the off-pulsed interval of the pulsar \gcygpsr. As a consequence of this approach, we are able to unveil, for the first time, the \gray emission from the SNR at energies lower 1 GeV. We showed that, if the emission underlying the intense \gray radiation from the pulsar is taken into account with an off-pulse analysis, a complex pattern of \gray emission emerges, showing the SNR-accelerated particles interacting with the surrounding gas clouds and ISR photons. We found that the \grid off-pulsed emission profile above 400 MeV appears to partially cover the SE rim of the shell. On the other hand, most of the GeV-TeV emission (\lat/VERITAS) is located in the NW side.

The \gray SED detected by AGILE from the whole SNR angular extent is presented. 

A lepton-dominated double population model can explain both the radio and the high-energy \gray emission from the whole extent of the SNR, indicating that the MeV-GeV emission is related to Bremsstrahlung processes in a high density medium, whereas the GeV-TeV radiation is due to both non-thermal Bremsstrahlung ($E_{\gamma} \lesssim 250$ GeV) and IC processes ($E_{\gamma} \gtrsim 250$ GeV) in a lower density medium. Observations strongly suggest that the first population of leptons is associated with the SE part of the shell (in coincidence with the bright radio feature DR4 and a higher gas density), and the second population with the NW rim.

The AGILE observations are contributing in a crucial way to elucidate the low-energy domain of the spectrally resolved \gray morphology of \gcyg.

Future observations at radio wavelengths with the Square Kilometre Array (SKA) and at TeV energies with the Cherenkov Telescope Array (CTA) will shed new light in the understanding of the physics behind the non-thermal emission from this SNR.

\acknowledgments
Research partially supported by the ASI grant no. I/028/12/5 and the INAF PRIN SKA-CTA: ``Towards the SKA and CTA: discovery, localization and physics of transients''.\\
The authors thank the anonymous referee for her/his stimulating comments on the manuscript. We also thank A. De Rosa (INAF-IAPS) for the useful discussions concerning the X-ray spectral data of the SNR, M. Giusti (INAF-IAPS) for technical support on the manuscript, and M. Lemoine-Goumard (Bordeaux University -- CNRS-IN2P3) for providing the \lat $TS$ map (previously published in \citealp{ackermann_17}), from which we extracted the contours in Fig.~\ref{gamma_cyg_agile_fermi}.

\software{AGILE Scientific Analysis Software (Build 25) \citep{bulgarelli_12,bulgarelli_19}, Pulsar (version 9) \citep{pellizzoni09}.}

\end{document}